\documentclass[aps,pra,epsfigure,twocolumn,showpacs]{revtex4}
\usepackage{dcolumn}    
\usepackage{bm} 
\usepackage{graphicx}
\usepackage{amsmath}    
\usepackage{latexsym}
\usepackage{amsfonts}   
\usepackage{amssymb}
\usepackage{array}      
\usepackage{epsfig}
\usepackage{txfonts}
\usepackage{float}
\usepackage{color}

\newcommand{\ket}[1]{\left\vert#1\right\rangle}

\newcommand{\bra}[1]{\left\langle#1\right\vert}

\begin{document}

\title{Effect of inter-particle interaction in a free oscillation atomic interferometer}

\author{Thom\'as Fogarty$^{1,2}$, Anthony Kiely$^2$, Steve Campbell$^{1,2}$ and Thomas Busch$^{1,2}$}

\affiliation{$^1$Quantum Systems Unit, OIST Graduate University, Okinawa 904-0495, Japan\\
$^2$Department of Physics, University College Cork, Republic of Ireland}

\begin{abstract}
We investigate the dynamics of two interacting bosons repeatedly scattering off a beam-splitter in a free oscillation atom interferometer. Using the inter-particle scattering length and the beam splitter probabilites as our control parameters, we show that even in a simple setup like this a wide range of strongly correlated quantum states can be created. This in particular includes the NOON state, which maximizes the quantum Fisher information and is a foremost state in quantum metrology.
\end{abstract}
\date{\today}
\maketitle

\section{Introduction}
\label{Introduction}

In all scientific pursuits accurate measurements are crucial and many of the most successful techniques in quantum metrology make use of the principles of interferometry. Recently significant progress has been made by recognizing that the use of quantum correlations, in particular entanglement, enables us to make the most precise measurements physically allowed by the Heisenberg uncertainty principle~\cite{nature,atomoptical,BECinter1,bana}.  Optical interferometers are able to generate a wide range of quantum correlated states, such as the NOON state~\cite{mitchell,walther,rarity,afek}, however a major draw back are their short coherence times. In order to enhance measurement precision through the use of entanglement, longer coherence times are highly desirable.

Making use of atomic ensembles can enhance the lifetime of a generated state, however the often unwanted and hard-to-control scattering interactions can make them  difficult to work with~\cite{atomoptical}. Nevertheless, some remarkable progress has been made using Bose-Einstein condensates (BECs) as resources~\cite{BECinter1,BECinter2, sackett08}. For example, in the presence of attractive interactions these allow for the formation of bright soliton states, which are non-dispersive and have been suggested as good candidates for the creation of macroscopic spatial superpositions \cite{streltsov,castin,gardiner,weiss12}. Furthermore, the ubiquitous presence of harmonic traps for ultracold atoms has led to new ideas for interferometry designs based on the periodic trap dynamics \cite{martin,zozulya,nakagawa,sackett}. Such schemes, which present a viable approach to atomic interferometry require often minimal experimental efforts and are referred to as free oscillation atom interferometers.

Here we investigate the behaviour of two ultracold atoms in such a free oscillation atom interferometer and fully take their mutual interaction into account. We start with two bosonic atoms located on one side of a harmonic trap split by a delta-potential, whose strength can be adjusted. The atoms are then released and allowed to scatter off the barrier twice, thus realizing a Michelson type interferometer. By employing numerical diagonalization techniques we are able to exactly solve the model and determine the atom pair's full density matrix at any moment in time. While previous studies have explored how different properties of the trap affect the performance of an interferometer~\cite{sackett,dunningham}, here we rigorously assess the effects which different interaction regimes and beam splitter ratios have on the non-classical nature of the states created. We quantify this by calculating the Quantum Fisher information (QFI)~\cite{braunstein}, and show that for a certain range of parameters this simple setup can generate the highly desirable NOON state.

The remainder of the paper is organized as follows. In Sec.~\ref{toolbox} we formalize the physical model and present the various tools to be used throughout. In Sec.~\ref{resultsattractive} we assess the case when the atoms possess an attractive interaction, while in Sec.~\ref{resultsreplusive} we explore the repulsive regime. The experimental feasibility is considered in Sec.~\ref{experiment} and in Sec.~\ref{conclusions} we present our conclusions and discussions of the results.

\section{Preliminaries}
\label{toolbox}
\subsection{The Model}
The atomic interferometer we consider is a harmonic trap punctuated centrally by a delta function potential. The delta function barrier will act as a beam splitter for the interacting atoms, and for simplicity we restrict our investigation to the case of two atoms. We assume the trap is such that only longitudinal motion is permitted and transverse motion is tightly restricted, thus forming a quasi one-dimensional system. The Hamiltonian is then given by
\begin{equation}
\label{hamiltoniank}
\begin{split}
\mathcal{H}_{\Omega}= & \sum^{2}_{n=1} \left( -\frac{\hbar^{2}}{2m}\frac{\partial^2}{\partial {x}_n^2}+\frac{1}{2} m \Omega^2 x_n^2 + \kappa_0 \delta(x_n) \right) 
              		  +V\left( | {x}_1-{x}_2 | \right),
\end{split}		  
\end{equation}
where $m$ is the mass of each particle, $\Omega$ the frequency of the harmonic potential and $\kappa_0$ is the height of the $\delta$-function barrier. Throughout the paper, unless otherwise stated, all units are dimensionless.  At low temperatures the boson-boson interaction, $V$, can be approximated by a point-like potential
\begin{equation}
\label{interaction}
V\left( | x_1 - x_2 | \right) = g_{1D}~\delta \left( | x_1-x_2 | \right),
\end{equation} 
where $g_{1D}$ is the one-dimensional coupling constant between particles defined in terms of the three-dimensional scattering length as $g_{1D}=\frac{4\hbar^2a_{3D}}{ma_{\perp}^2(1-\frac{Ca_{3D}}{a_{\perp}})}$ with $C\simeq1.4603$ and $a_{\perp}=\sqrt{\frac{\hbar}{\mu \omega_\perp}}$ with $\omega_\perp$ the transverse trap frequency and $\mu=m/2$ the reduced mass \cite{Olshanni}. This parameter will be central in our analysis of different regimes and can be experimentally tuned by applying a Feshbach resonance, a powerful technique that is well established in cold atomic physics \cite{julienne}. 

\begin{figure}[t]
  {\bf (a)}\hskip3.5cm{\bf (b)}
  \includegraphics[width=\linewidth]{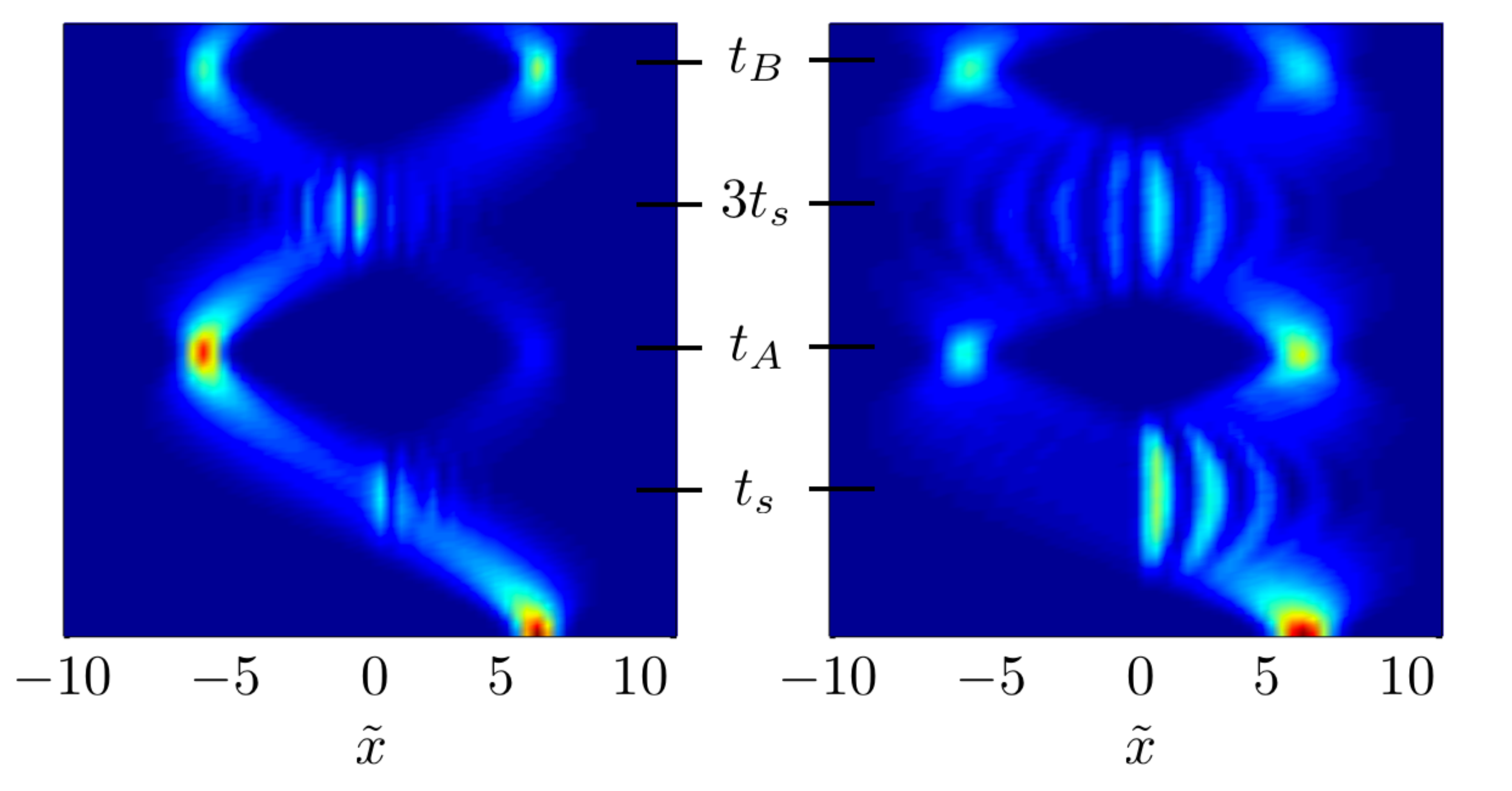}    
  \caption{(Color Online). Single particle density versus time for {\bf (a)} two attractive and {\bf (b)} two repulsive atoms. The barrier is positioned at $\tilde{x}=0$ and the two particles are initially trapped at $\tilde d=6$ with $\epsilon=5.164$. At time $t_s$ the particles scatter off the barrier and come to rest at time $t_A$ at the classical turning point. At time $3t_s$ the atoms recombine and scatter a second time and come to rest again at time $t_B$.}
\label{figspd}
\end{figure}

Initially the two atoms are prepared in a separate tight harmonic trap a distance $d$ from the centre of the interferometer trap and their state is given by the ground state of the Hamiltonian
\begin{equation}
\label{hamiltonian0}
\mathcal{H}_{\omega}=\sum^{2}_{n=1} \left( -\frac{\hbar^{2}}{2m}\frac{\partial^2}{\partial x_n^2}+\frac{1}{2} m \omega^2 (x_n-d)^2 \right) + g_{1D}~\delta\left( | x_1-x_2 | \right).
\end{equation}
Here $\omega$ is the trap frequency of the preparatory trap. In the following we will make use of natural units such that the coordinates are rescaled with respect to the characteristic scales of the harmonic oscillator, $\tilde{x}_n=x_n/a_{\Omega}$ and  $\tilde{E}_n=E_n/(\hbar\Omega)$. Here $a_{\Omega}=\sqrt{\hbar/m\Omega}$ is the width of the trap ground state in the axial direction. Thus we have
\begin{equation}
\label{hamiltoniank02}
\begin{split}
\tilde{\mathcal{H}}_{\Omega} =& \sum^{2}_{n=1} \left( -\frac{1}{2}\frac{\partial^2}{\partial \tilde{x}_n^2}+\frac{1}{2} \tilde{x}_n^2 + \kappa \delta\left(\tilde{x}_n\right) \right)
				    +g~\delta \left( | \tilde{x}_1-\tilde{x}_2 | \right),
\end{split}
\end{equation}

\begin{equation}
\label{hamiltoniank01}
\tilde{\mathcal{H}}_{\omega} = \sum^{2}_{n=1} \left(-\frac{1}{2}\frac{\partial^2}{\partial \tilde{x}_n^2}+\frac{1}{2}\epsilon^2 \left(\tilde{x}_n-\tilde{d}\right)^2 \right) +g~\delta \left( | \tilde{x}_1-\tilde{x}_2 | \right),
\end{equation}
where $\epsilon=\omega/\Omega$ is the ratio of the preparatory trap frequency to the one of the interferometer trap, $g=g_{1D} /(a_{\Omega}\hbar \Omega)$, $\tilde{d}=d/a_{\Omega}$ and $\kappa=\kappa_0 /(a_{\Omega} \hbar\Omega)$. 

In order to solve the Hamiltonians, $\tilde{\mathcal{H}}_{\Omega}$ and $\tilde{\mathcal{H}}_{\omega}$, we must determine the single particle eigenstates and associated energies. For the preparatory stage the time-independent Schr\"odinger equation 
\begin{equation}
\tilde{\mathcal{H}}_{\omega}\psi(\tilde{x_1},\tilde{x_2})=\tilde{E}_{n}^{'}\psi_n(\tilde{x_1},\tilde{x_2}) 
\end{equation}
can be treated by taking advantage of the separability of the Hamiltonian into centre of mass and relative coordinates, for which the solutions are well known \cite{bus98}. However, the Schr\"odinger equation for the interferometer potential $\tilde{\mathcal{H}}_{\Omega}\phi_{n}(\tilde{x_1},\tilde{x_2})=\tilde{E}_{n}\phi(\tilde{x_1},\tilde{x_2})$ does not allow such a luxury and must be solved numerically using, for example, a discrete variable representation (DVR) method~\cite{DVR1,DVR2}. The DVR method allows exact diagonalization of the many body Hamiltonian and scales as ${N_{p}}^{N^2}$, where $N_{p}$ is the number of points taken in configuration space.  In general this is numerically expensive, however the restriction of our analysis to $N=2$ particles allows the calculations to be tractable. Time evolution is then achieved by constructing the time dependent wave function in terms of the eigenstates of the hamiltonian $\tilde{\mathcal{H}}_{\Omega}$
\begin{equation}
\psi_m(\tilde{x}_1,\tilde{x}_2,t)=\sum_{n=0}^{\infty}a_{mn}\phi_n(\tilde{x}_1,\tilde{x}_2)e^{-i\tilde{E}_{n}t}
\end{equation}
in which 
\begin{equation}
a_{mn}=\int \psi_m(\tilde{x}_1,\tilde{x}_2) \phi_n(\tilde{x}_1,\tilde{x}_2) d\tilde{x}_1d\tilde{x}_2
\end{equation}
is the overlap of the individual solutions of the two Hamiltonians. 
Due to the atom's initial potential energy they will gain velocity, scatter at the barrier at time $t_s=\pi/2\Omega_\delta$ ({\it scattering A})
and  return to the classical turning points of the trap at $t_A=\pi/\Omega_{\delta}$ (see the dynamics of the single particle density in Fig.~\ref{figspd}). Here $\Omega_{\delta}\leq \Omega$ is an effective trap frequency adjusted to the presence of the delta function barrier.  At time $3t_s=3\pi/2\Omega_{\delta}$ the atoms scatter a second time ({\it scattering B}) and again return to the classical turning points at $t_B=2\pi/\Omega_\delta$. This setup resembles an atomic Michelson interferometer. While the following analysis can easily be performed by describing the barrier with a well localised potential of any shape, our choice of a delta-function is done to clearly isolate the interesting physical effects and does not constitute any loss of generality. A delta-function potential is a good approximation to a localised laser potential or an interaction with an atomic impurity fixed at $x=0$. In the first case the barrier height $\kappa$ can be experimentally tuned by changing the laser intensity, whereas in the second case a Feshbach resonance can be employed. This, coupled with the capacity to alter the inter-particle interaction, means we have a highly adaptable system with which to create superposition states.

\begin{figure*}[t]
  {\bf (a)}\hskip3.5cm{\bf (b)}\hskip3.5cm{\bf (c)}
\psfig{figure=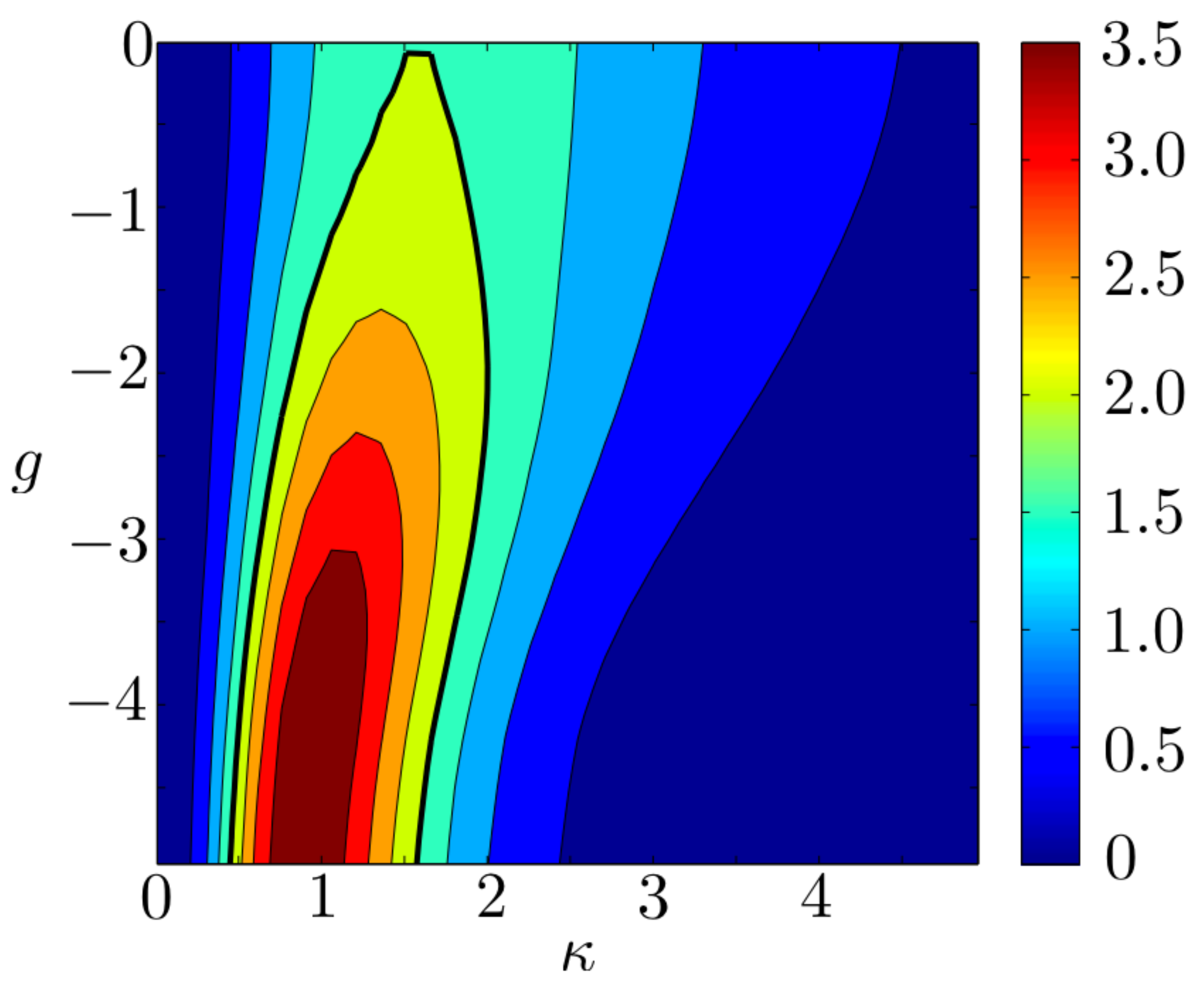,width=5.2cm}~~~~\psfig{figure=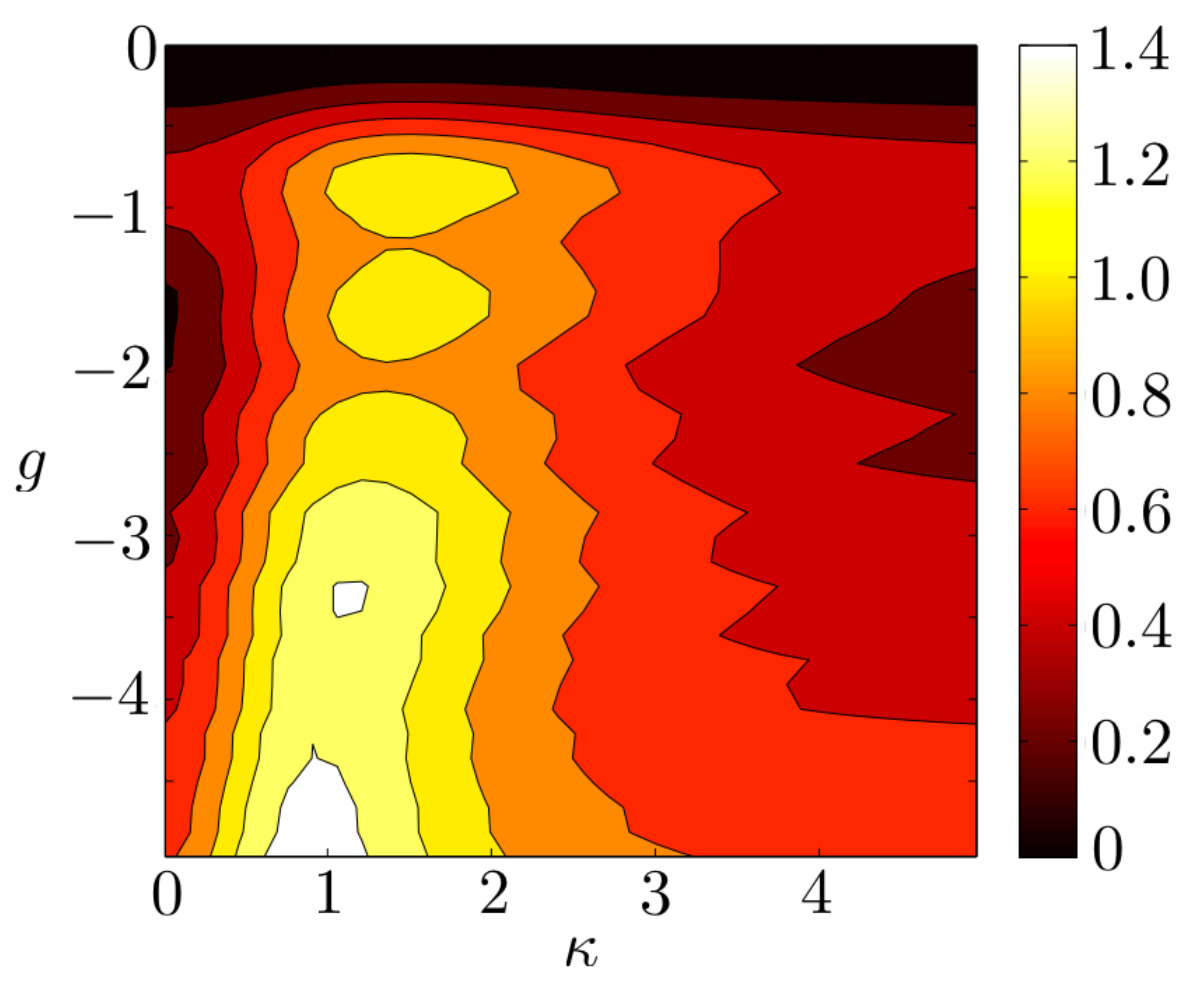,width=5.2cm}~~~~\psfig{figure=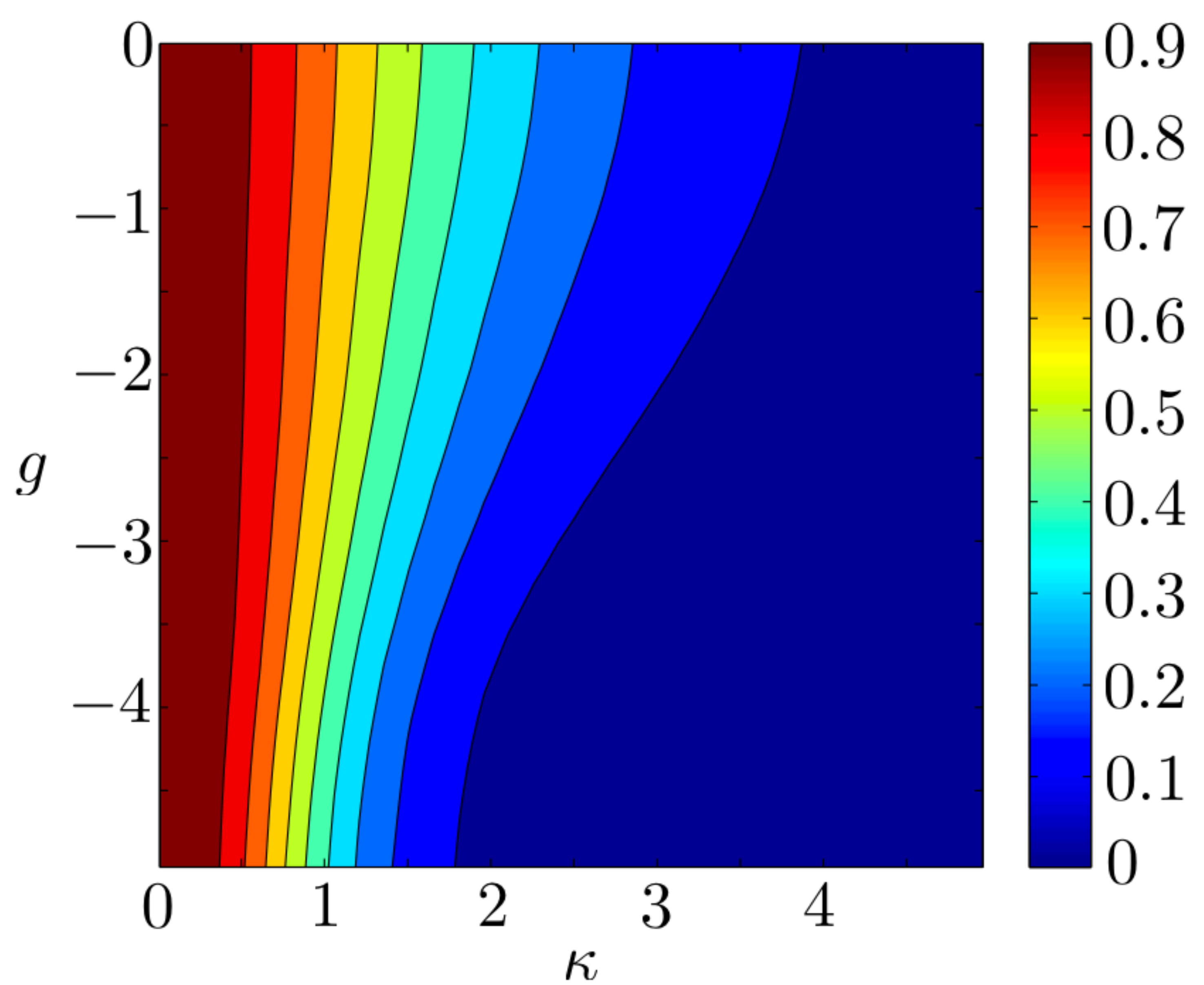,width=5.2cm}\\
{\bf (d)}\hskip3.5cm{\bf (e)}\hskip3.5cm{\bf (f)}
\psfig{figure=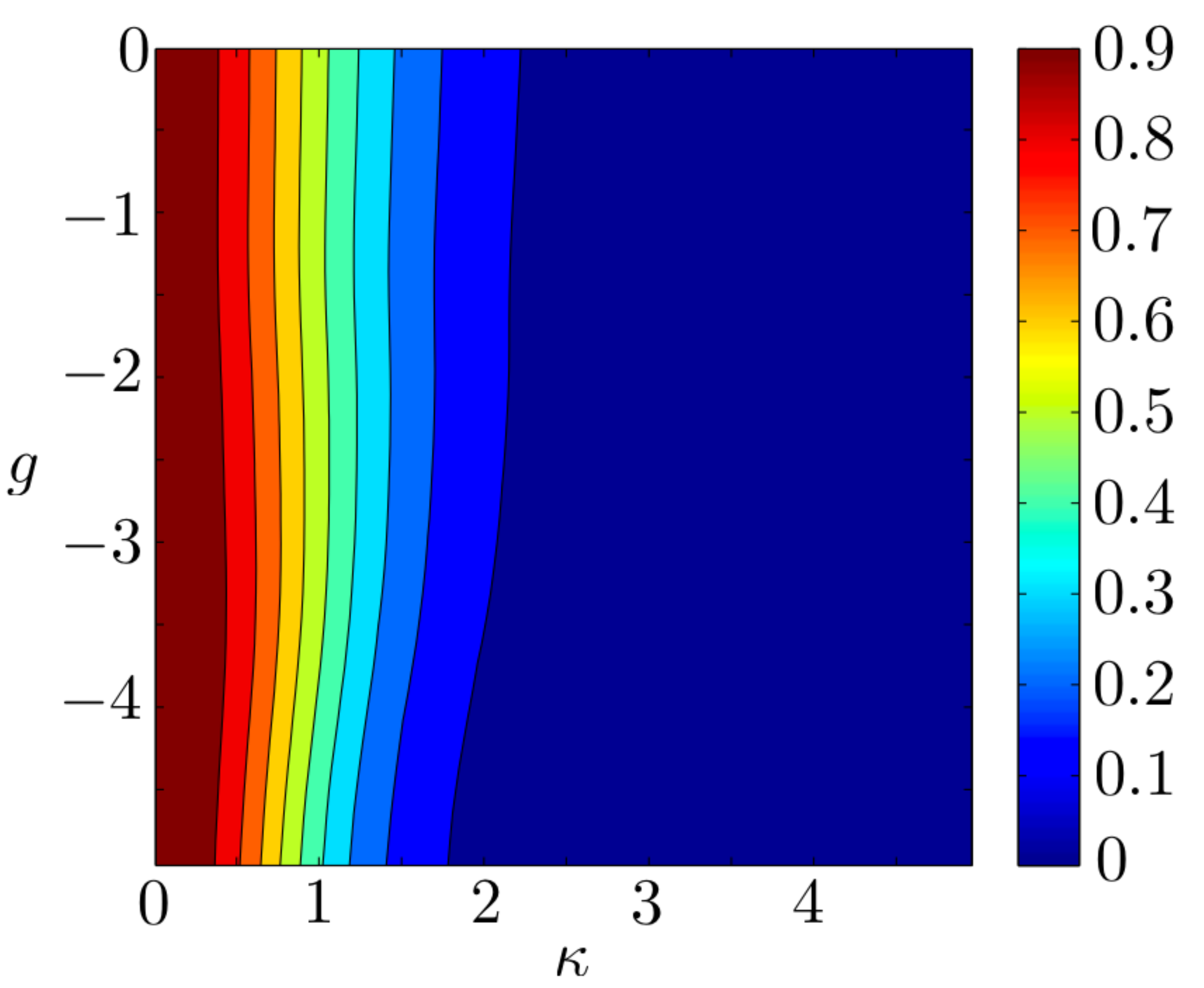,width=5.2cm}~~~~\psfig{figure=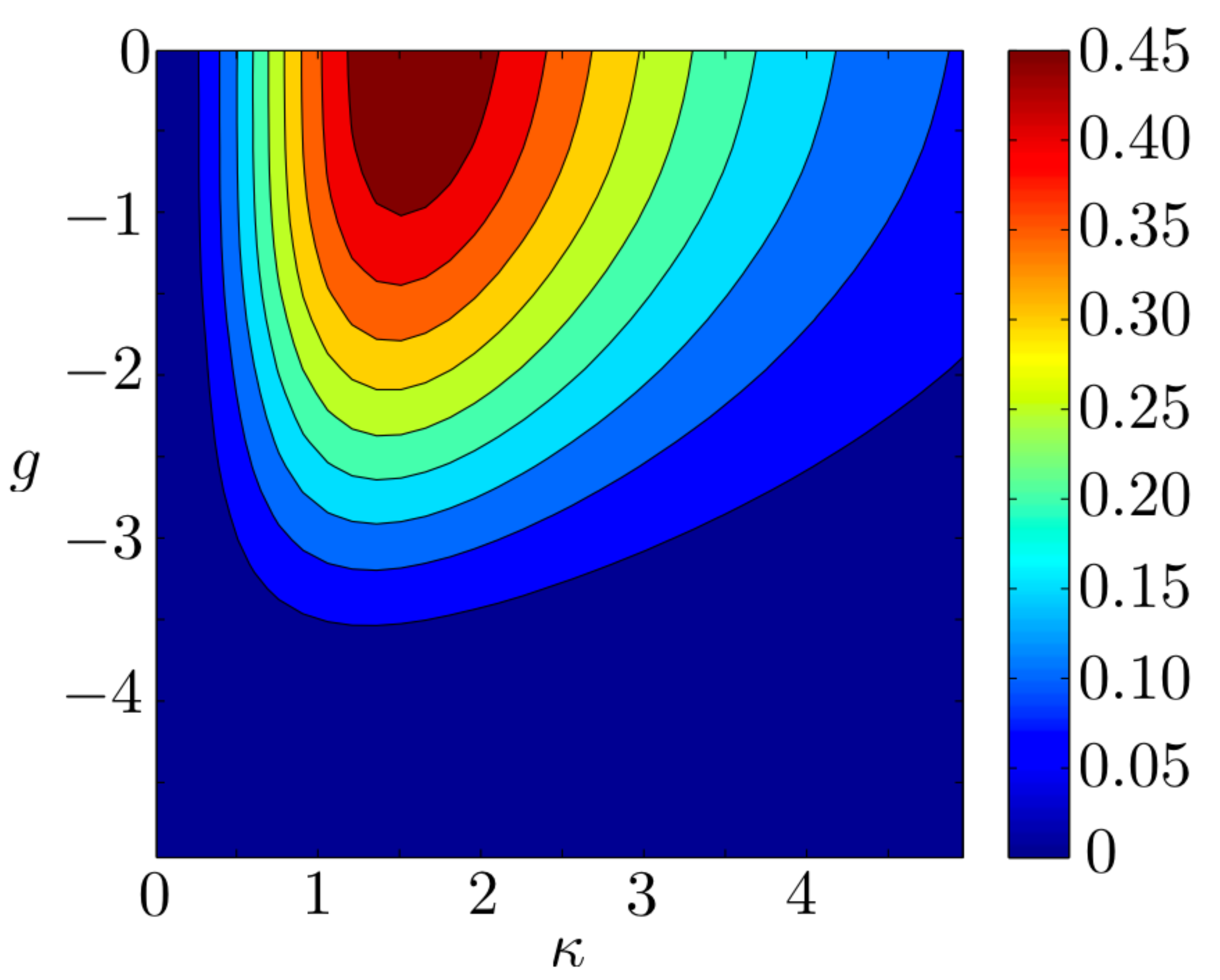,width=5.2cm}~~~~\psfig{figure=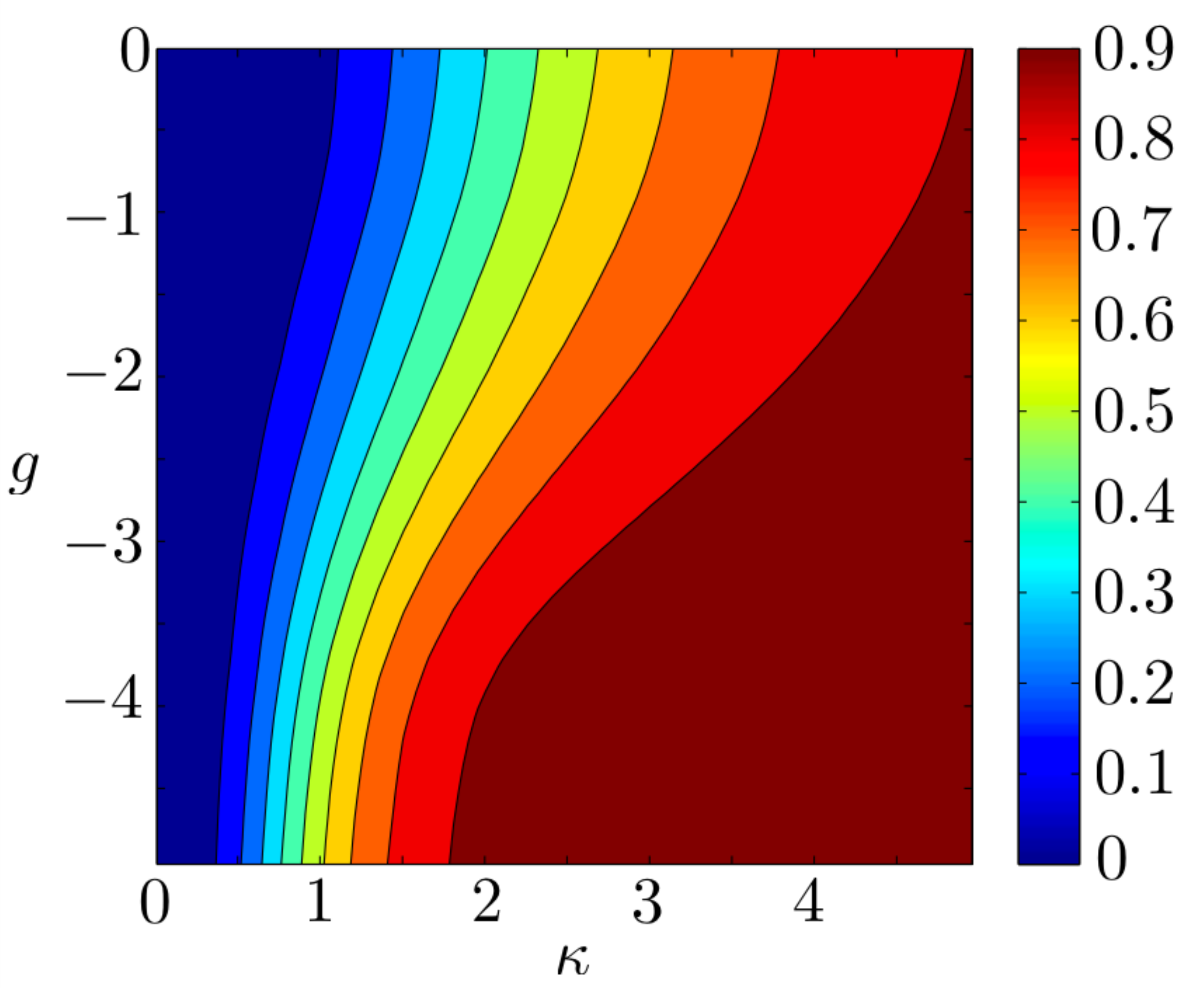,width=5.2cm}
\caption{(Color Online). Contour plots for the {\bf (a)} QFI, {\bf (b)} vNE, {\bf (c)} transmission coefficient ($T$), and population coefficients for states {\bf (d)} $\ket{20}$, {\bf (e)} $\ket{11}$ and {\bf (f)} $\ket{02}$ at time $t_A$ as a function of the attractive interaction strength $g$ and the barrier height $\kappa$ for $\epsilon=5.164$ and $\tilde d=6$.}
\label{attractivet1}
\end{figure*}

\subsection{Quantum Fisher Information and von Neumann Entropy}
In the following we will thoroughly explore the features of the states at the times $t_A$ and $t_B$.
For this we will use the Quantum Fisher Information \cite{fisher}, which defines the amount of information an observable can obtain about an unknown parameter. A state with maximal QFI will allow the most sensitive measurements, for example of relative phases. For a pure state $\ket{\psi(\varphi)}$ the QFI is defined as
\begin{equation}
\mathcal{F}_Q=4\left( \bra{\psi'(\varphi)} \psi'(\varphi)\big> - \vert \bra{\psi'(\varphi)} \psi(\varphi)\big> \vert^2 \right),
\end{equation}
where $\ket{\psi'(\varphi)}=\partial \ket{\psi(\varphi)}/\partial \varphi$. For separable states the maximum QFI is equal to the number of particles (or quanta), $N$, used in the interferometer, which corresponds to the shot-noise limit. However one can go beyond this limit by using entangled states which can yield a maximum QFI of $N^2$, the Heisenberg limit \cite{dunningham}. One particularly important class of states that reach this limit are the so-called NOON states,
\begin{equation}
\ket{\psi_{NOON}}=\frac{1}{\sqrt{2}}\left( \ket{N}\ket{0}+\ket{0}\ket{N} \right),
\end{equation}
which in our scheme corresponds to both particles being simultaneously on the left-hand side (LHS) and on the right-hand side (RHS) of the barrier. Thus we are we are looking at the spatial correlations of the two atoms \cite{gooldheaney}. Of course, NOON states are not the only interesting non-classical states to study in interferometry, however as we are examining $N\!=\!2$ they are the most prominent and our study is in line with the optical counterparts recent state of the art experiments~\cite{2photon}. For larger systems, i.e. $N\!>\!2$ we expect the scheme to be extremely versatile.

The scheme generates a pure bi-partite entangled state, and as such we will also make use of the von Neumann entropy (vNE) to quantify the entanglement of the atoms. It is defined by the entropy of the reduced single particle density matrix, $\rho$, as
\begin{equation}
S(\rho)=\text{Tr}[\rho~\text{log}_2~\rho] =\sum_i \lambda_i \text{log}_2 \lambda_i,
\end{equation}
where the $\lambda_i$ are the eigenvalues of this matrix and defined by $\int \rho(\tilde{x}_1,\tilde{x}_2)\chi_i(\tilde{x}_2)d\tilde{x}_2=\lambda_i\chi_i(\tilde{x}_1)$. Due to the required symmetry of the wave-function for identical bosons one must be careful when dealing with the vNE as an entanglement measure in certain situations as discussed in~\cite{gooldtonks}. In our setup, however, the dynamical scattering process and constant interaction between the particles ensures that any finite von Neumann entropy signals genuine entanglement. Since the vNE measures the total entanglement, and therefore accounts for both inter-particle and spatial entanglement, it can be expected to show some qualitative differences to the QFI.

\begin{figure*}[t]
{\bf (a)}\hskip3.5cm{\bf (b)}\hskip3.5cm{\bf (c)}
\psfig{figure=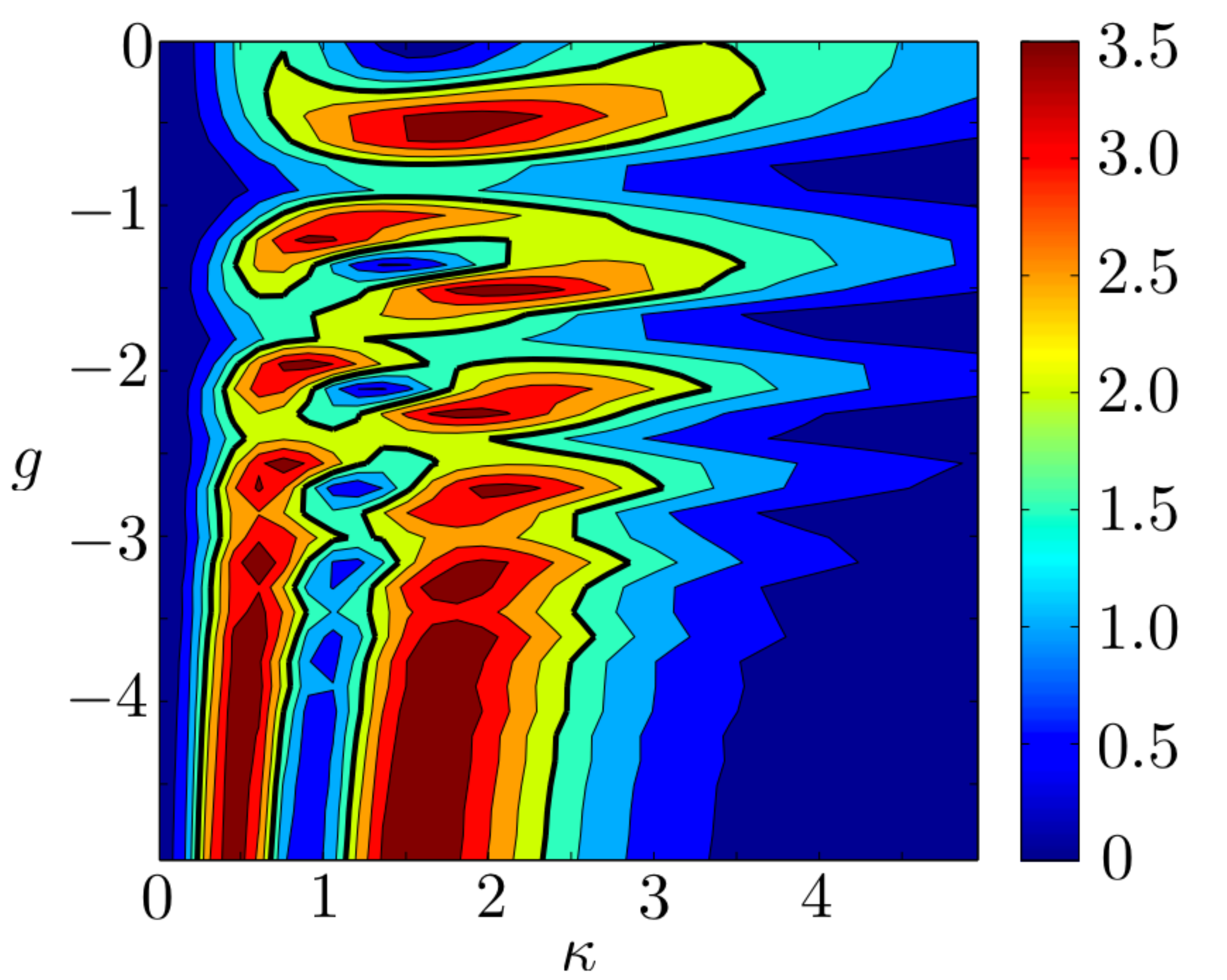,width=5.2cm}~~~~\psfig{figure=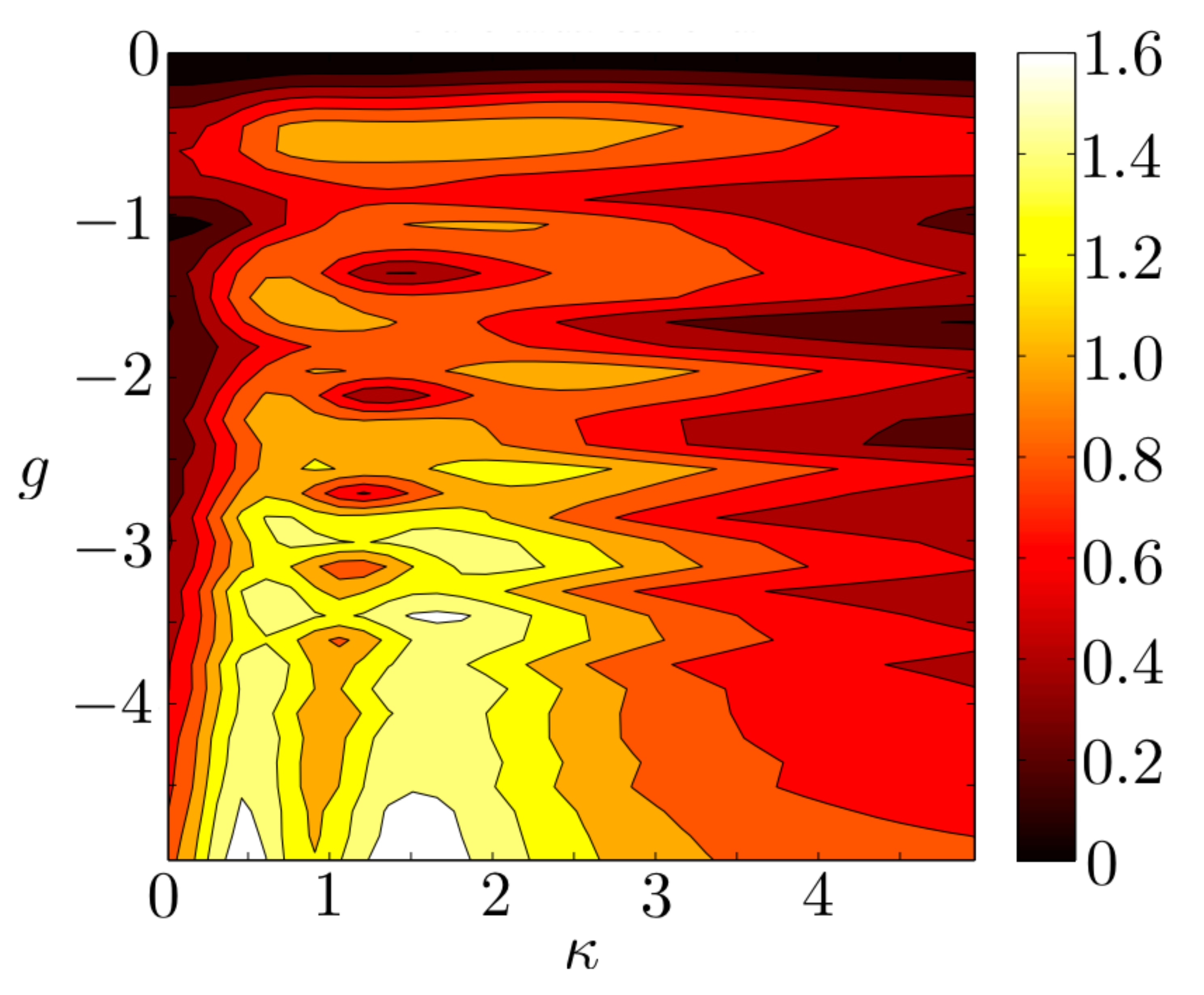,width=5.2cm}~~~~\psfig{figure=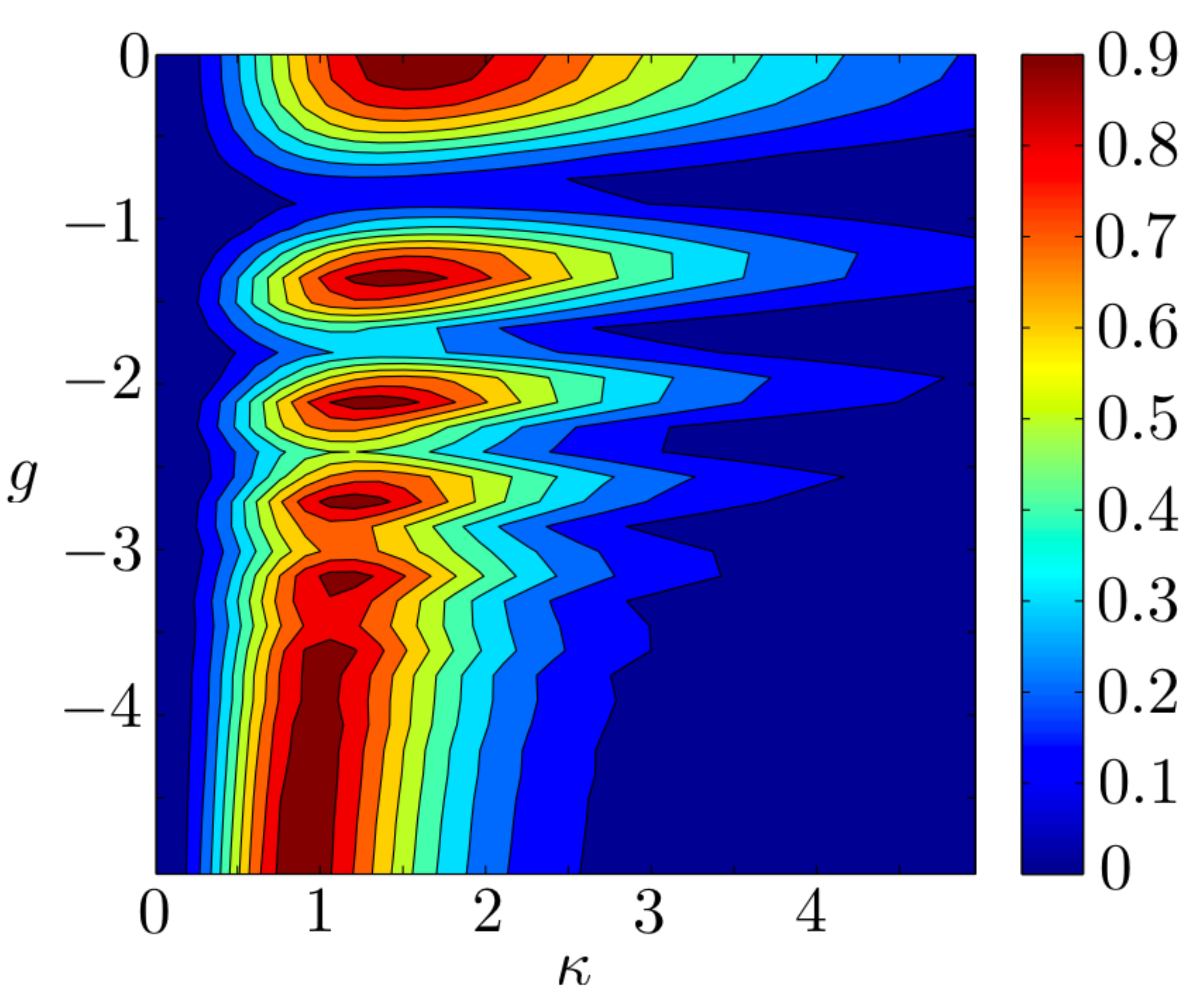,width=5.2cm}\\
{\bf (d)}\hskip3.5cm{\bf (e)}\hskip3.5cm{\bf (f)}
\psfig{figure=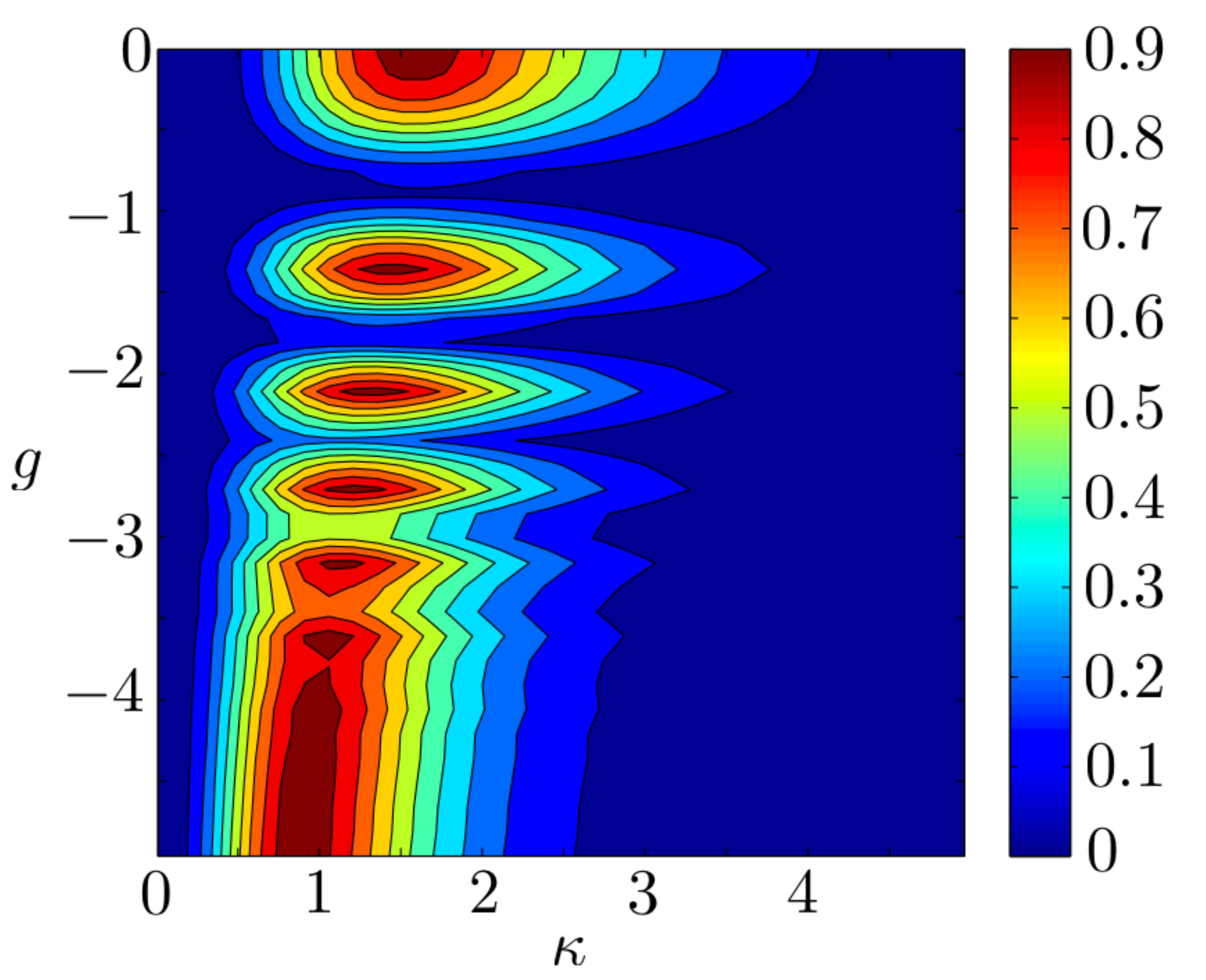,width=5.2cm}~~~~\psfig{figure=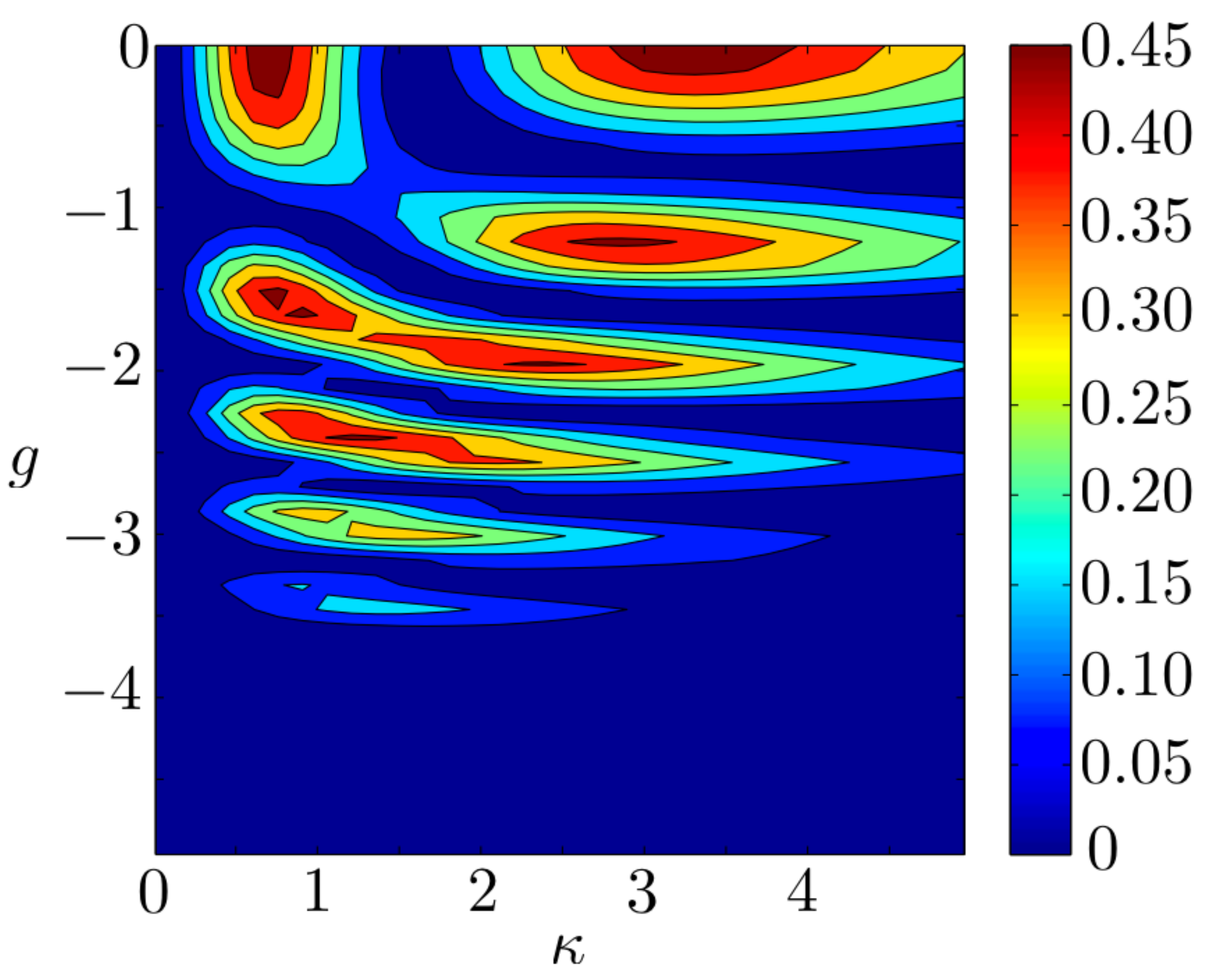,width=5.2cm}~~~~\psfig{figure=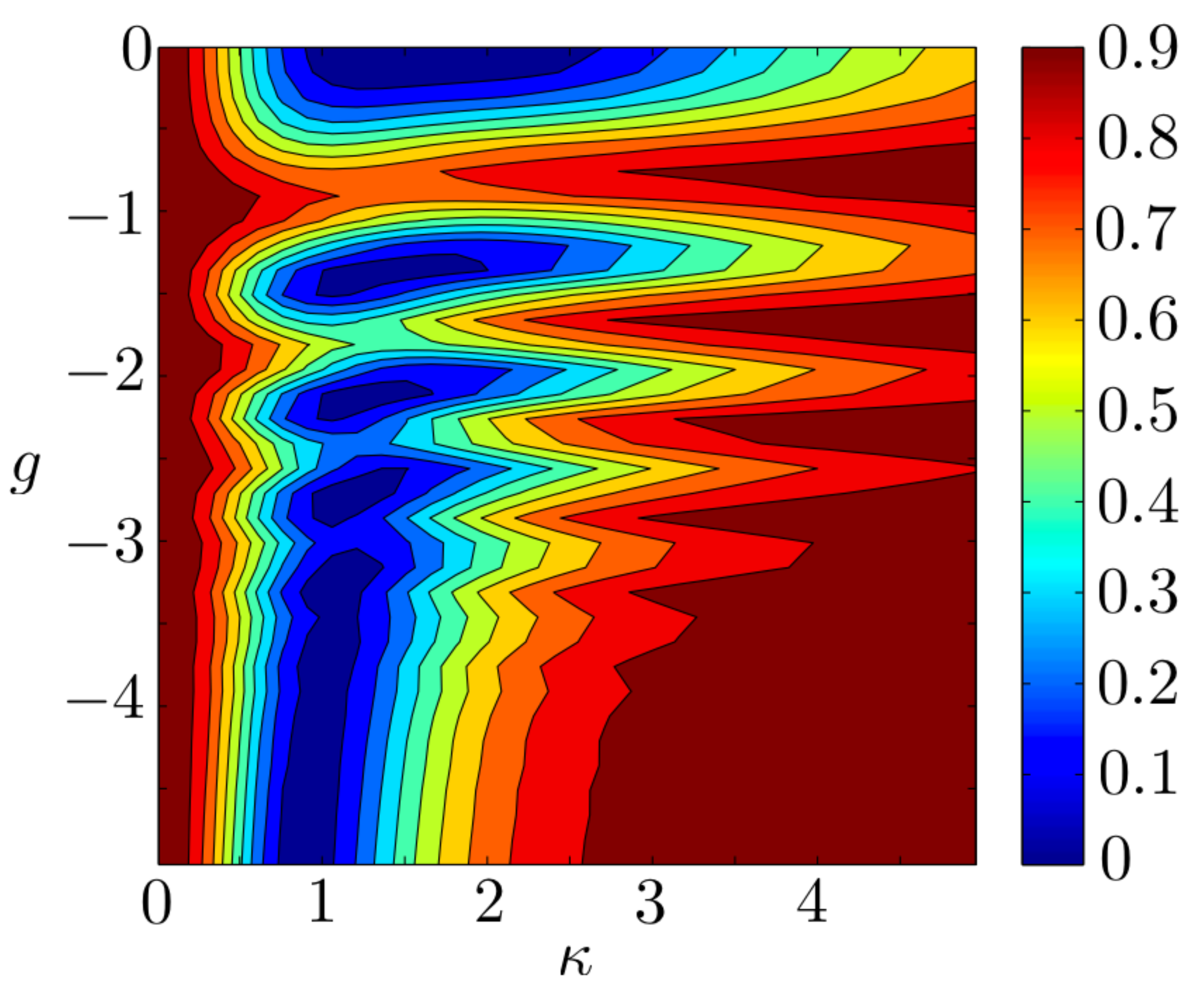,width=5.2cm}
\caption{(Color Online). Contour plots for the {\bf (a)} QFI, {\bf (b)} vNE, {\bf (c)} transmission coefficient ($T$), and population coefficients for states {\bf (d)} $\ket{20}$, {\bf (e)} $\ket{11}$, and {\bf (f)} $\ket{02}$  as a function of attractive interaction strength $g$ and barrier height $\kappa$ at time $t_B$ for the same initial state as in Fig.~\ref{attractivet1}.}
\label{attractivet2}
\end{figure*}

\section{Analysis of different interaction regimes}
\subsection{Attractive Interactions}
\label{resultsattractive}
\subsubsection{Scattering A}
We first examine the state of an attractive dimer after scattering once off the delta barrier. In Fig.~\ref{attractivet1} {\bf (a)} we plot the QFI as a function of attractive interaction strength, $g$, and barrier height $\kappa$. The thick black line signifies the classical shot noise limit at $\mathcal{F}_Q=N=2$, which is attainable for separable states. Interestingly we find that even for a weakly interacting dimer we can exceed this bound. As we increase the attractive interactions between the atoms the QFI increases to its maximal obtainable value of $N^2=4$ for a barrier height of $\kappa\approx1$. In panel {\bf (b)} we see that the behavior of the vNE is qualitatively in agreement, although more complex. The small scale details are due to the inter-particle interaction leading to a constantly varying inter particle entanglement, which is not captured in the calculation of the QFI. Looking at the transmission coefficient $T$, Fig.~\ref{attractivet1} {\bf (c)}, we see that the maximum QFI is achieved for symmetric splitting ($T=0.5$). To confirm the state generated in this situation is the NOON state $\frac{1}{\sqrt{2}}(\ket{20}+\ket{02})$, we show the various population coefficients in Figs.~\ref{attractivet1} {\bf (d)-(f)}. One can immediately see that the region in which the QFI is maximized corresponds to states for which the $\ket{11}$ component is suppressed and the $\ket{20}$ and $\ket{02}$ components are equally populated. This can intuitively be understood by realising that the relatively strong attractive interaction within the dimer makes it hard to split the pair of atoms into one on the left and one on the right. In fact, the situation is analogous to the one of bright, atomic solitons, where it has been shown that macroscopic superposition states can be created by moving an atomic soliton through a barrier of finite width \cite{streltsov,weiss12}.

\subsubsection{Scattering B}
After the second scattering process the dynamics becomes more complex for the attractive dimer. Examining the QFI, Fig.~\ref{attractivet2} {\bf (a)}, we see that even for weakly attractive particles we can attain $\mathcal{F}_Q\approx4$ and as we increase the interaction strength we find the QFI peak at two values of the barrier height, $\kappa$. The behavior of the vNE, panel {\bf (b)}, is qualitatively similar and is also mirrored in the transmission coefficient, $T$, panel {\bf (c)}. Once again, for $T=0.5$ we see a maximum QFI. The most striking feature is clearly the intricate series of maxima appearing in all panels. This is due to the phases accumulated by the atoms at the beam splitter and when traveling along its two arms. For the case of non-symmetric splitting the different interaction energies of the particles lead to a difference in phase, which in turn leads to the observed interference fringes. We see the same qualitative behavior in the various population coefficients shown in Fig.~\ref{attractivet2} {\bf (d)-(f)}, where the maximum QFI again corresponds to a suppression of the $\ket{11}$ state and an equal population of the other two states. Interestingly the value of $\kappa$ which resulted in a maximum QFI for {\it scattering A} results in a minimum QFI for {\it scattering B} for the same value of the interaction strength. 

\begin{figure}[b]
{\bf (a)}\hskip3.5cm{\bf (b)}
\psfig{figure=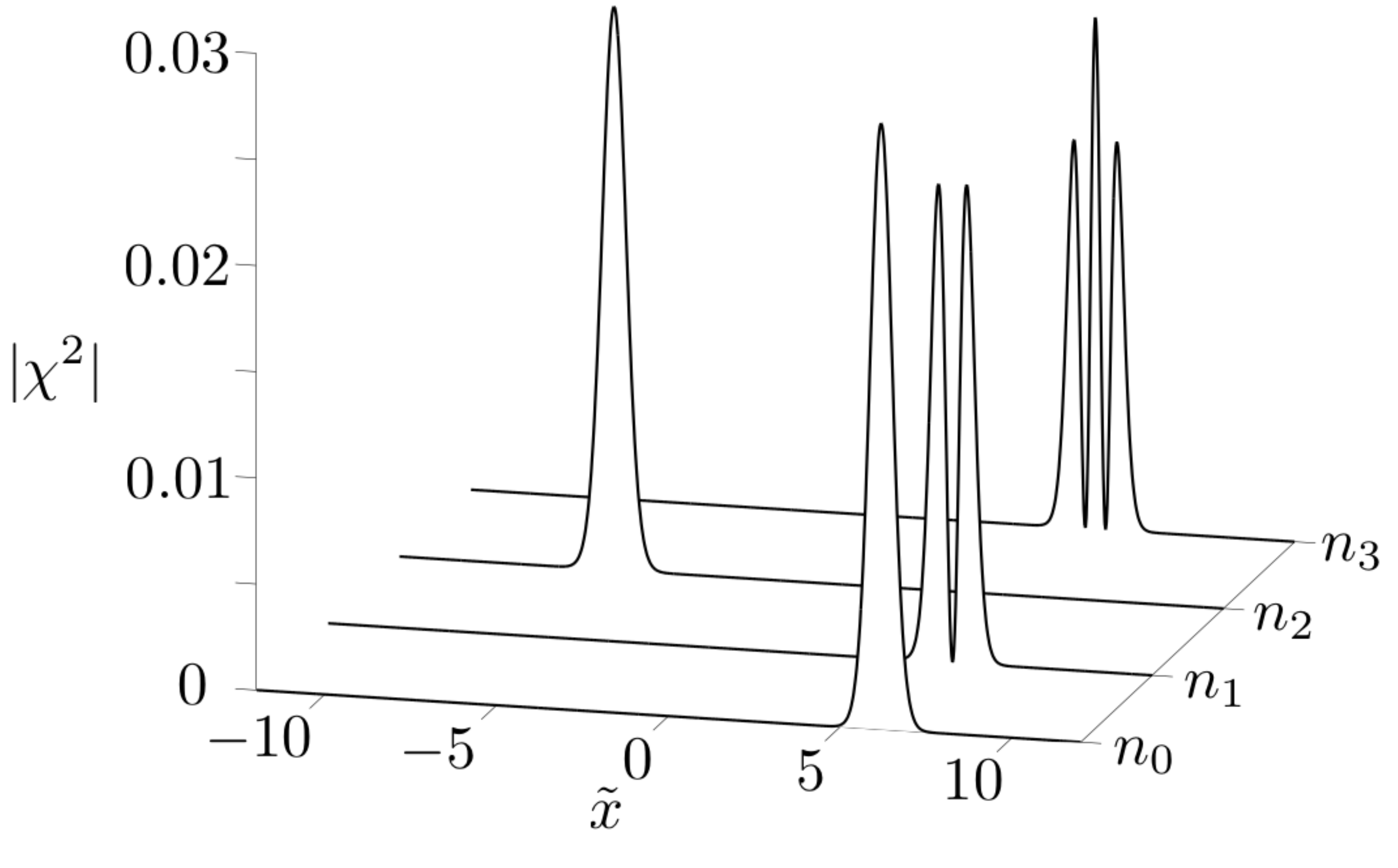,width=4cm}~~\psfig{figure=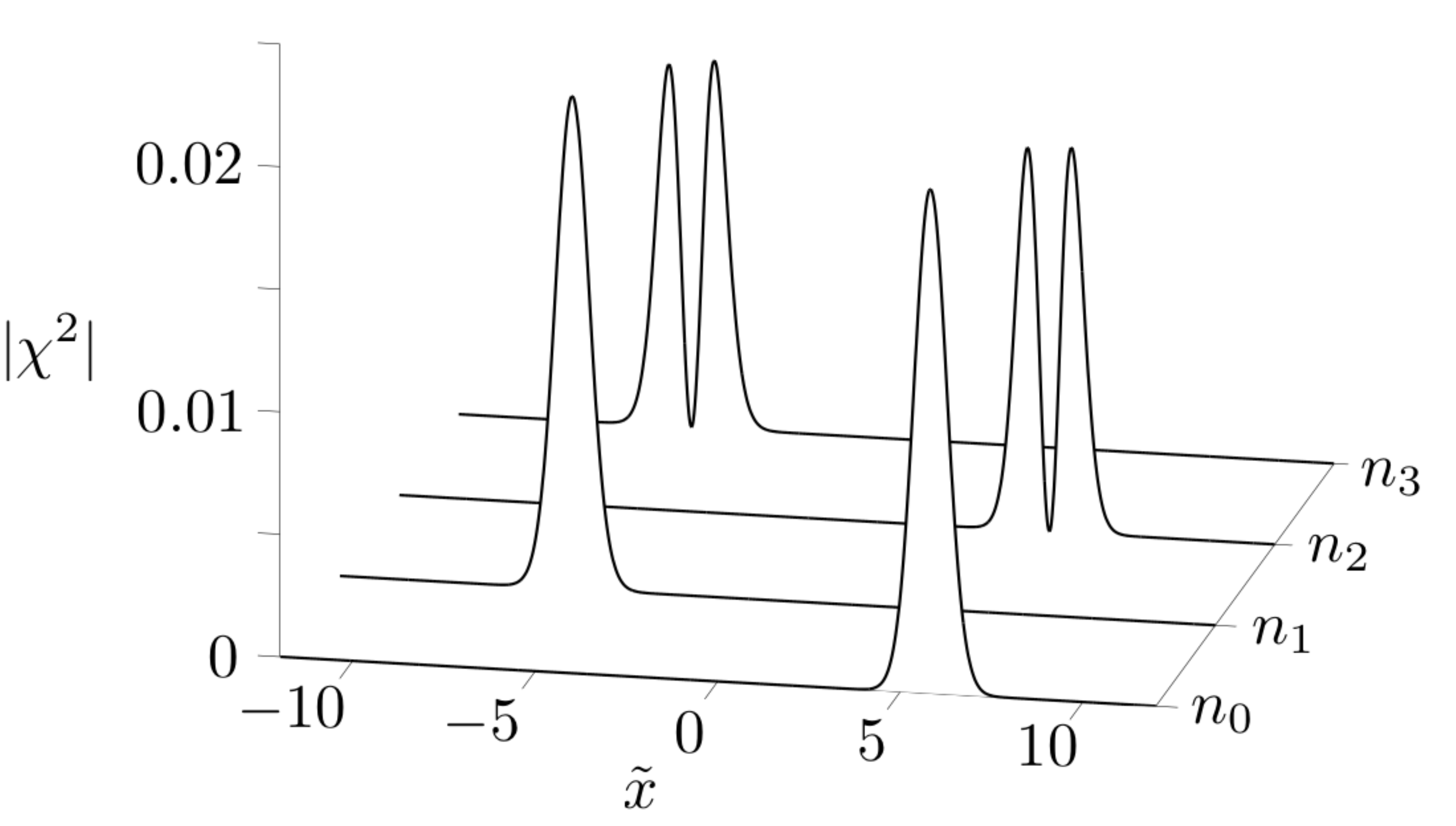,width=4cm}
\caption{Densities of four lowest lying eigenstates of the reduced single particle density matrix for the attractive dimer at times {\bf (a)} $t_A$ and {\bf (b)} $t_B$ for $\kappa=0.4$ and $g=-7$. At $t_A$ the lowest, the second excited and the third excited orbitals 
occupy the RHS of the trap and the first excited orbital is localised on the LHS. At $t_B$ the ground state and the second excited orbitals are on the RHS, whereas the other two orbitals are on the LHS.}
\label{attNOONorbitals}
\end{figure}

\begin{figure}[t]
{\bf (a)}\hskip3.5cm{\bf (b)}
\psfig{figure=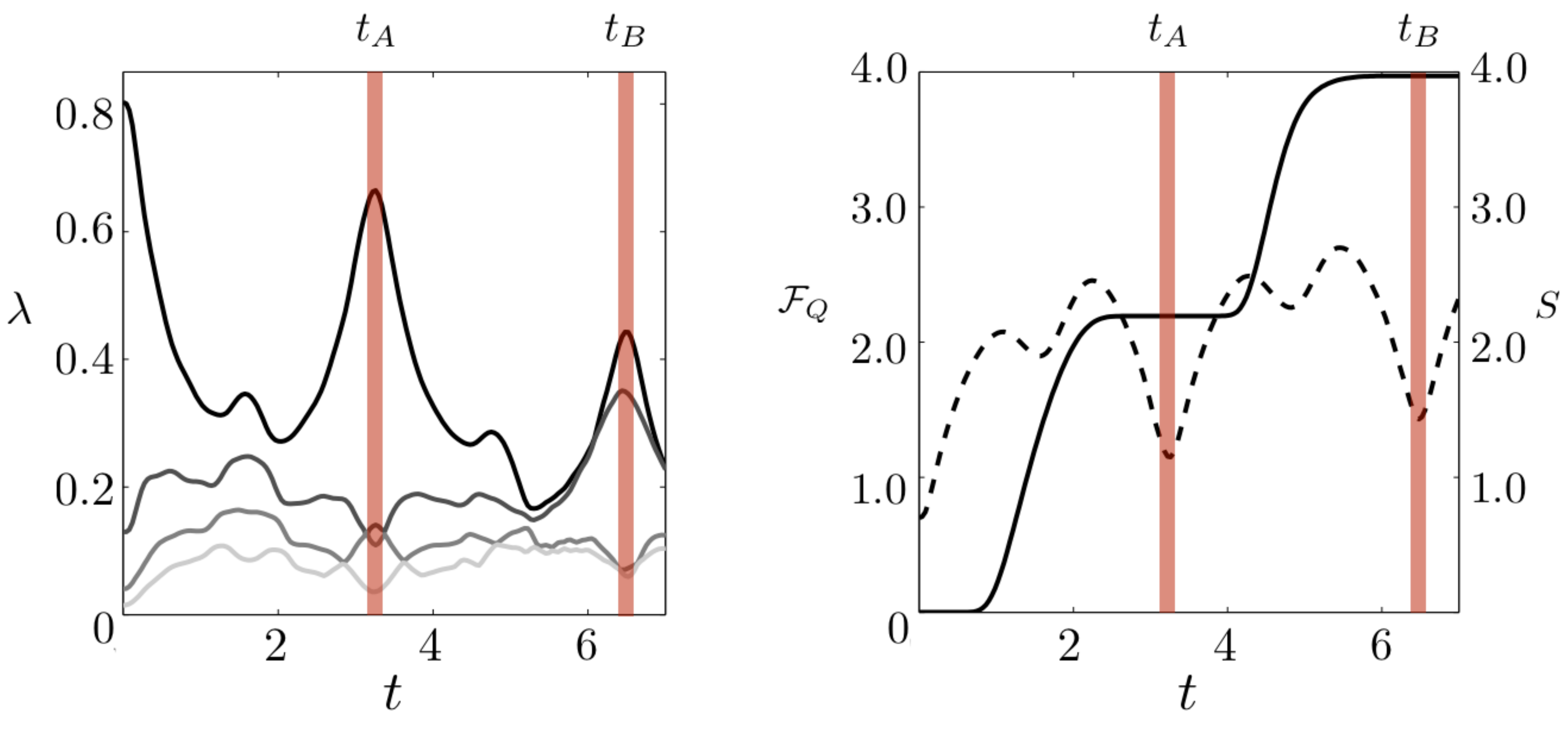,width=8.5cm}
\caption{{\bf (a)} The orbital occupation numbers versus time for attractively interacting particles $\kappa=0.4$ and $g=-7$. The darkest line corresponding to lowest energy orbital, with each progressively lighter shade representing the next higher energy orbital. {\bf (b)} The QFI (solid) and vNE (dashed) versus time. The maximum $F_Q$ reached is $N^2$ indicating that a NOON state is created.}
\label{attNOONFIvNE}
\end{figure}

To get a deeper insight into the process, we show in Fig.~\ref{attNOONorbitals} the densities of the four lowest lying eigenstates, $|\chi^2_i|$, ($i=0,1,2,3$), of the reduced single particle density matrix at times $t_A$ and $t_B$ and for $g=-7$ and $\kappa=0.4$. Fig.~\ref{attNOONorbitals} {\bf (a)}  shows that at $t_A$ (where we have $F_Q=2.035$) three of the orbitals are located on the RHS of the trap and one on the LHS.
This is a result of the large attractive interaction, which does not allow the particles to become spatially split by the barrier and thus the transmission coefficient is either $T=0$ or $T=1$. Even though each orbital occupies one side of the trap only, this is not a NOON state as can be seen by looking at the orbital occupation numbers (see Fig.\ref{attNOONFIvNE} {\bf (a)}). We find that at this point the lowest orbital has still the largest occupation number and higher lying ones have lower occupation. Fig.~\ref{attNOONorbitals} {\bf (b)} shows the situation at $t_B$ and we find two orbitals occupying each side of the trap. The orbital occupation probabilities for the ground and the first excited state are degenerate after the scattering event {\it B}, which proves the NOON nature of the state and explains the resulting $\mathcal{F}_Q=3.9998\approx N^2$ (see Fig.\ref{attNOONFIvNE} {\bf (b)}). The fact that the occupations are not exactly degenerate close to $t_B$ is due to the inter-particle interaction, which is reflected in the dynamics of the vNE, dashed line Fig.\ref{attNOONFIvNE} {\bf (b)}. It displays two pronounced dips exactly at $t_A$ and $t_B$, indicating a prominent change in the internal structure (due to the re-focussing at the classical turning point). Making the choice $\omega=\Omega$ leads to perfect degeneracy for all times after scattering B (and before the next scattering event).  Note that the step-wise behavior of the QFI is due to its sensitivity to only spatial correlations and it therefore only changes during the scattering process, while the constant interaction between the atoms gives rise to the varying vNE.

\subsection{Repulsive Interactions}
\label{resultsreplusive}
\subsubsection{Scattering A}
\begin{figure}[t]
{\bf (a)}\hskip3.5cm{\bf (b)}\\
\psfig{figure=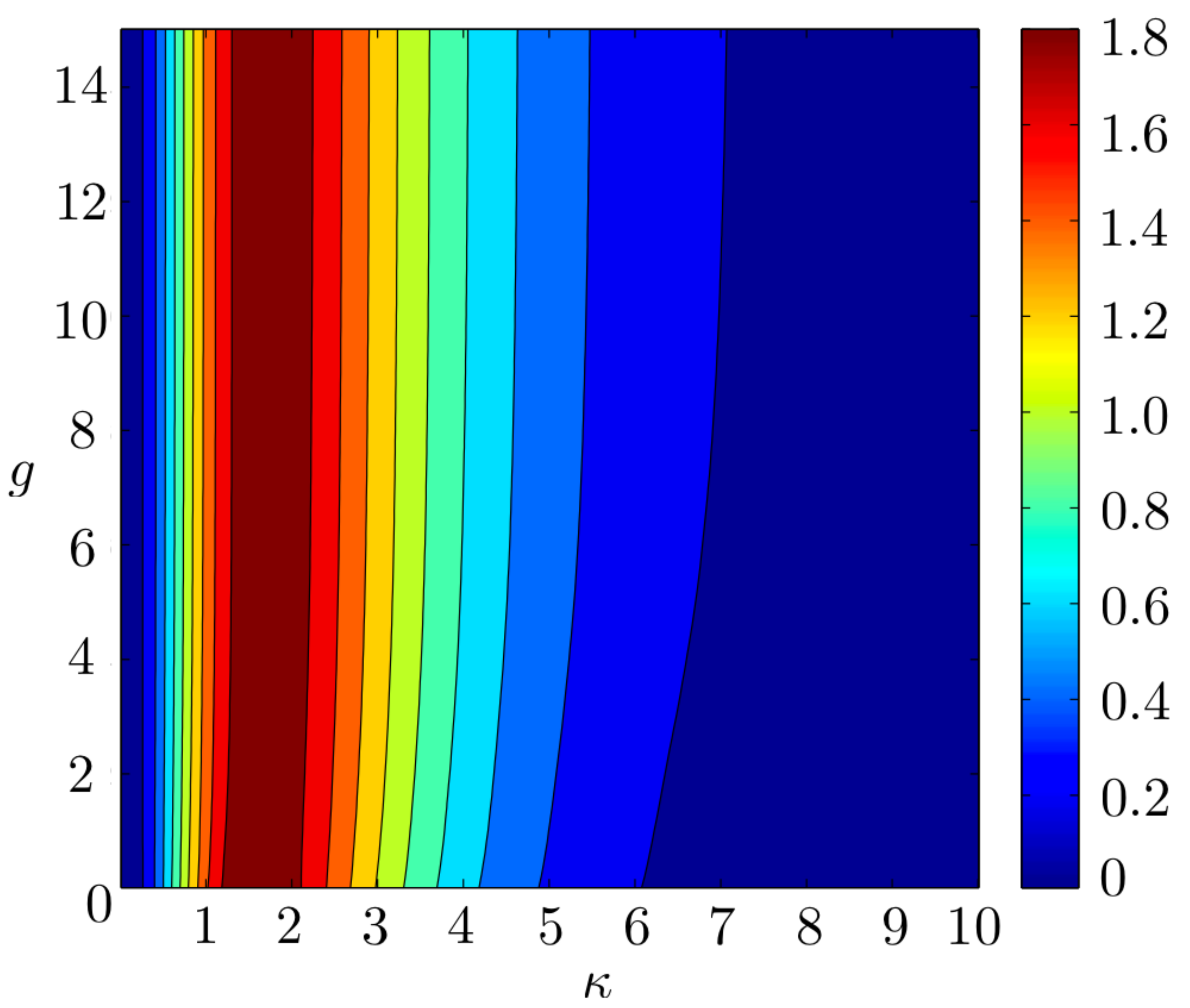,width=4.1cm}~~\psfig{figure=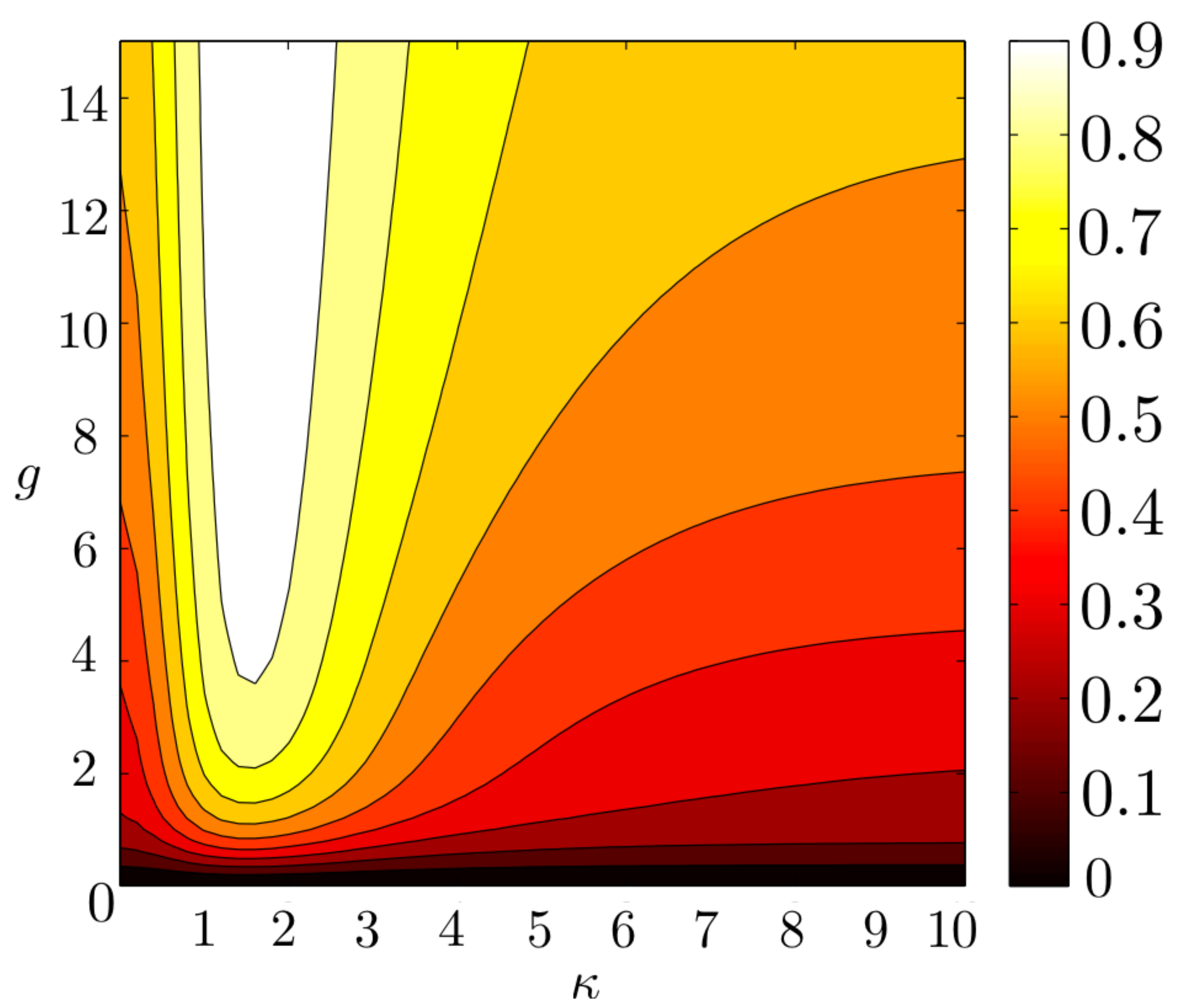,width=4.1cm}\\
{\bf (c)}\hskip3.5cm{\bf (d)}\\
\psfig{figure=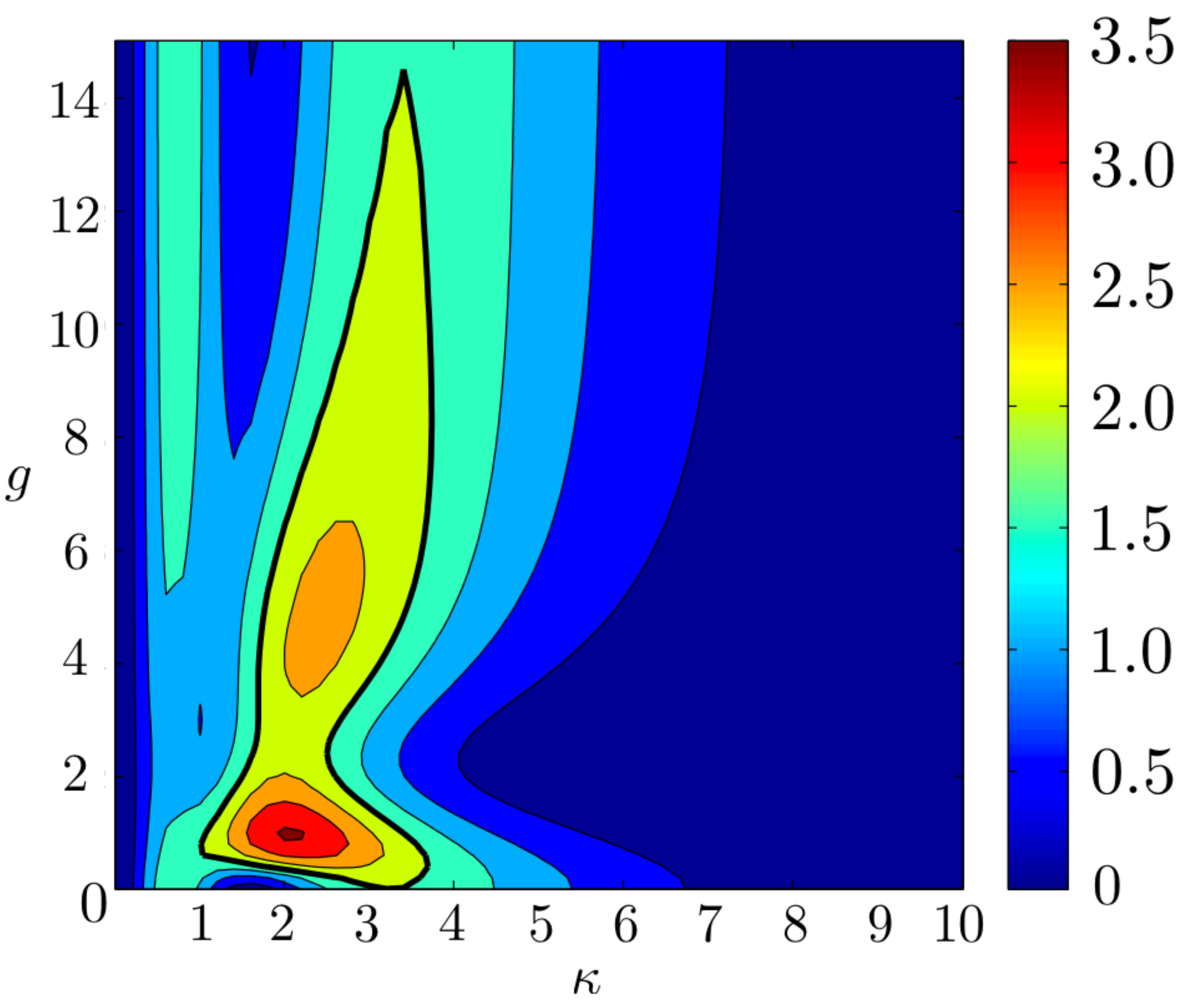,width=4.1cm}~~\psfig{figure=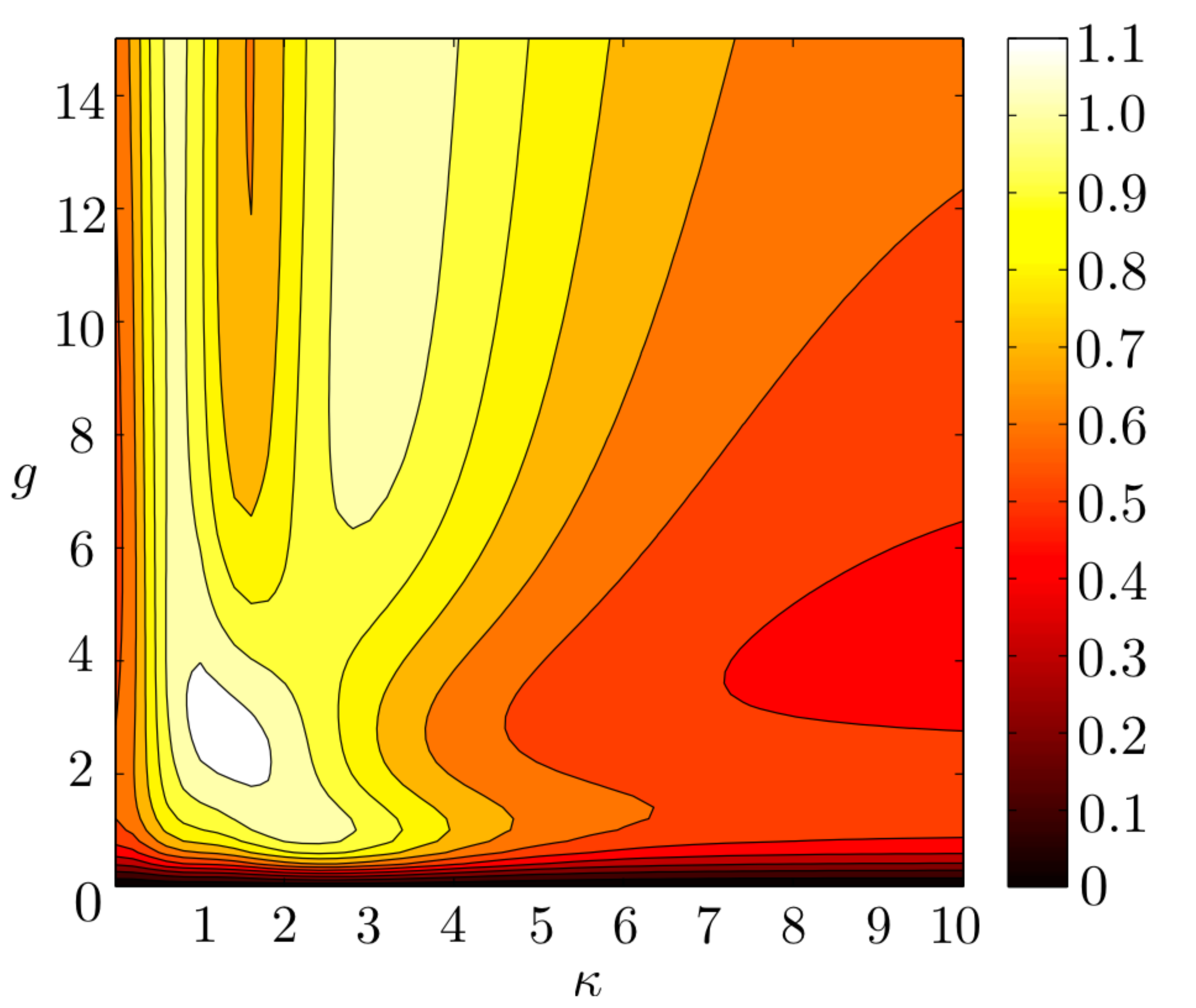,width=4.1cm}
\caption{(Color Online). Contour plots for the {\bf (a)} QFI and {\bf (b)} vNE at time $t_A$ after scattering once off the barrier and the {\bf (c)} QFI and {\bf (d)} vNE at time $t_B$ after scattering twice off the barrier, as a function of repulsive interaction strength $g$ and barrier height $\kappa$. The initial state is the same as in the attractive case discussed above.}
\label{repulsivet2}
\end{figure}

We now turn our attention to the case of repulsive interaction between the atoms. This regime gives rise to behaviors that do not promote the generation of spatial entanglement easily as the repulsive nature prefers a situation in which one atom occupies each side of the trap. Fig.~\ref{repulsivet2} {\bf (a)} shows that at time $t_A$ this set-up cannot produce states that outperform the best classically attainable states (since $\mathcal{F}_Q\leq2$) for the whole range of parameter space considered. The maximum $\mathcal{F}_Q=2$ occurs for a barrier height $\kappa=1.51$ regardless of the interaction strength $g$, reaching the classical limit for a transmission coefficient of $T=0.5$ (not shown). The vNE (Fig.~\ref{repulsivet2} {\bf (b)}) is also maximized for $T=0.5$, reaching 0.9 for $g=4$ and increasing to a maximum of approximately 1 for strongly repulsive atoms.

\begin{figure}[t]
\psfig{figure=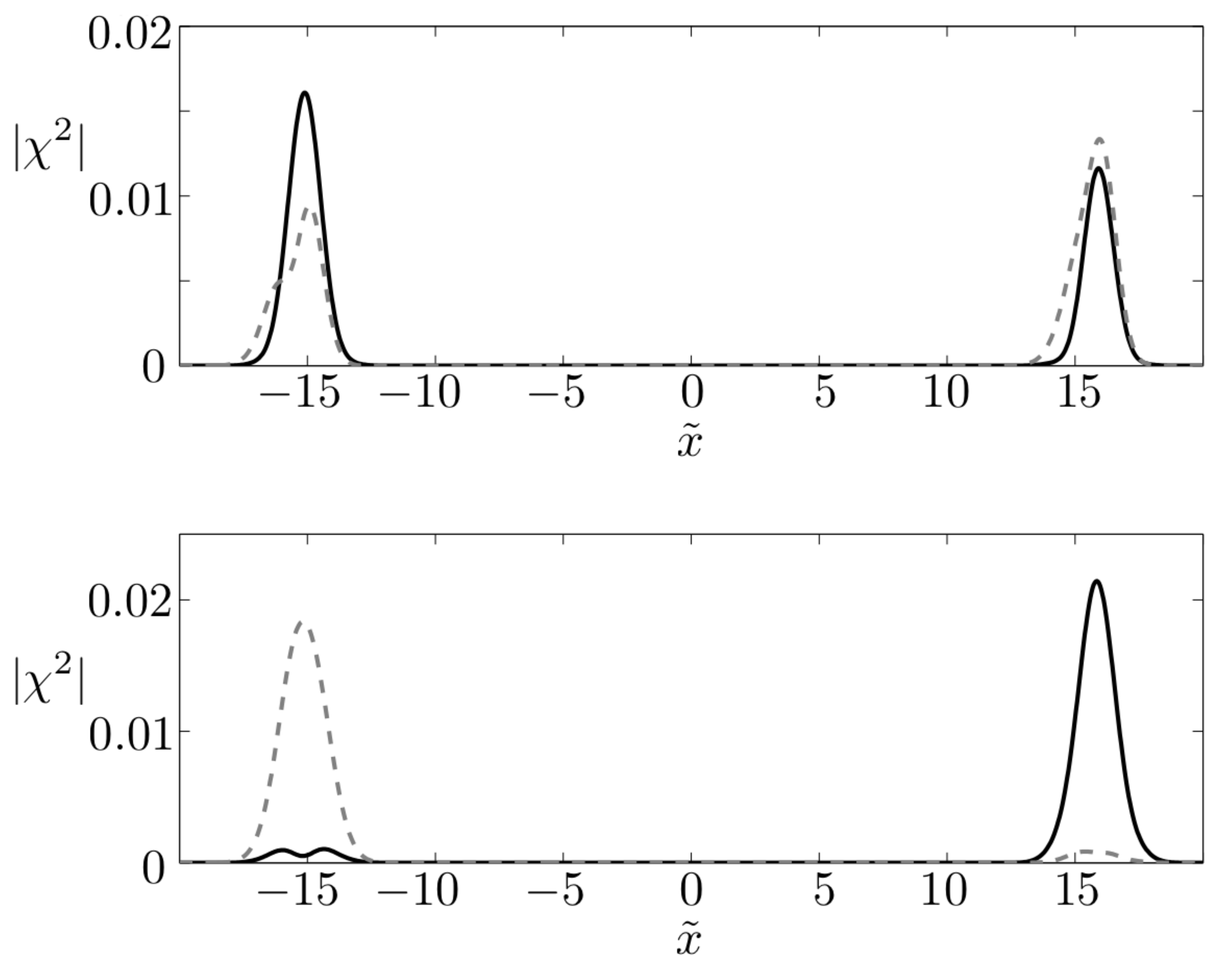,width=8cm}
\caption{The densities of the two lowest atomic orbitals of the reduced single particle density matrix at times $t_A$ [top panel] and $t_B$ [bottom panel] with a repulsive interaction with $g=1$ and $\kappa=2.1$. At $t_A$ the each orbital has equal probability to be in the LHS or RHS of the trap. At $t_B$ each orbital is at opposite sides of the trap, indicating a highly entangled quasi-NOON state.}
\label{fig1}
\end{figure}

\subsubsection{Scattering B}
Similar to the case of attractive interactions, the variety of states created after the second scattering process becomes much richer due to the phase acquired by the various components of the two-particle state. States with a $\mathcal{F}_Q>2$ can now be generated, however they are restricted to a much smaller area of the parameter space compared to the attractive interaction.  In Fig.~\ref{repulsivet2} {\bf (c)} we see for that small repulsive interaction a QFI of $\mathcal{F}_Q>3.5$ can be reached and the vNE in Fig.~\ref{repulsivet2} {\bf (d)} shows qualitatively similar behavior.
As the interaction $g$ is increased, the atoms enter the Tonks-Girardeau (TG) regime and the QFI approaches its classical limit of $2$, corresponding to the state $\ket{\psi}=\frac{1}{2}\ket{20}+\frac{1}{\sqrt{2}}\ket{11}+\frac{1}{2}\ket{02}$ resulting from a 50/50 splitting.  

\begin{figure}[t]
{\bf (a)}\hskip3.5cm{\bf (b)}
\psfig{figure=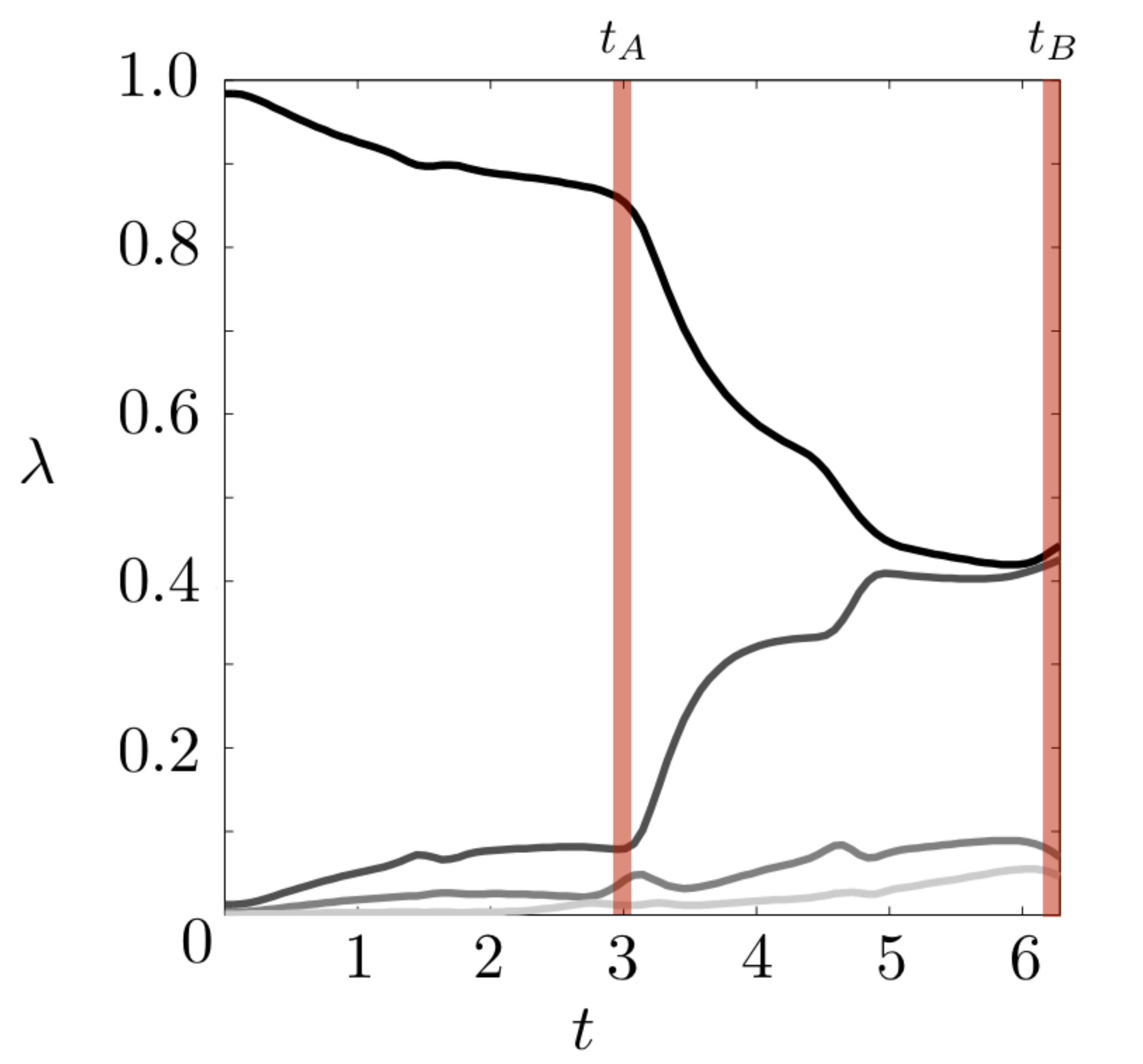,width=4cm}~~\psfig{figure=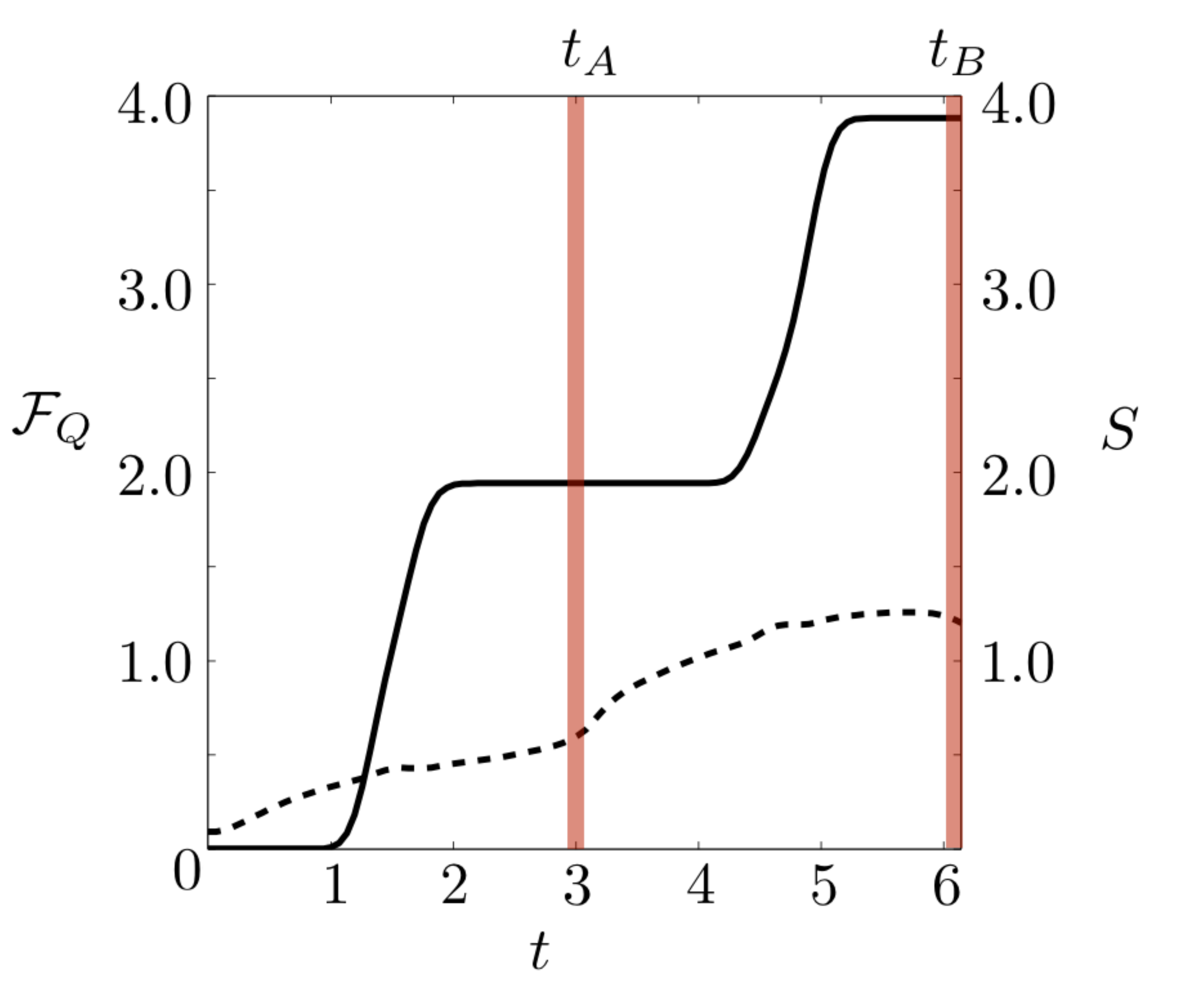,width=4.6cm}
\caption{{\bf (a)} The orbital occupation numbers are plotted versus time for a repulsive interaction with $g=1$ and $\kappa=2.1$. The darkest line corresponding to lowest energy orbital, with each progressively lighter shade representing the next higher energy orbital. At time $t_B$ the ground and first excited orbitals become nearly degenerate, while the higher lying ones are only weakly populated. {\bf (b)} The behavior of the QFI (solid) and vNE (dashed) as a function of time. The maximum $F_Q$ reached is $3.883$.}
\label{fig2}
\end{figure}

The state with the maximum QFI in this regime is achieved for $g=1$ and $\kappa=2.1$ and we show the corresponding lowest two eigenstates of the reduced single particle density matrix in Fig.~\ref{fig1}. At $t_A$ (upper panel) each orbital occupies both sides of the trap with nearly equal probability due to the $50/50$ splitting of the barrier and corresponds to $F_Q\approx 2$. At time $t_B$ (lower panel) each orbital almost fully localizes on one side of the trap and the respective occupation numbers approach double degeneracy, cf. Fig.~\ref{fig2} {\bf (a)}, indicating the presence of a superposition state in accordance with $\mathcal{F}_Q=3.883$, Fig.~\ref{fig2} {\bf (b)}. Due to the relatively weak repulsive interaction strength we see the vNE grow monotonically.

\section{Experimental Realization}
\label{experiment}
\begin{figure}[t]
\psfig{figure=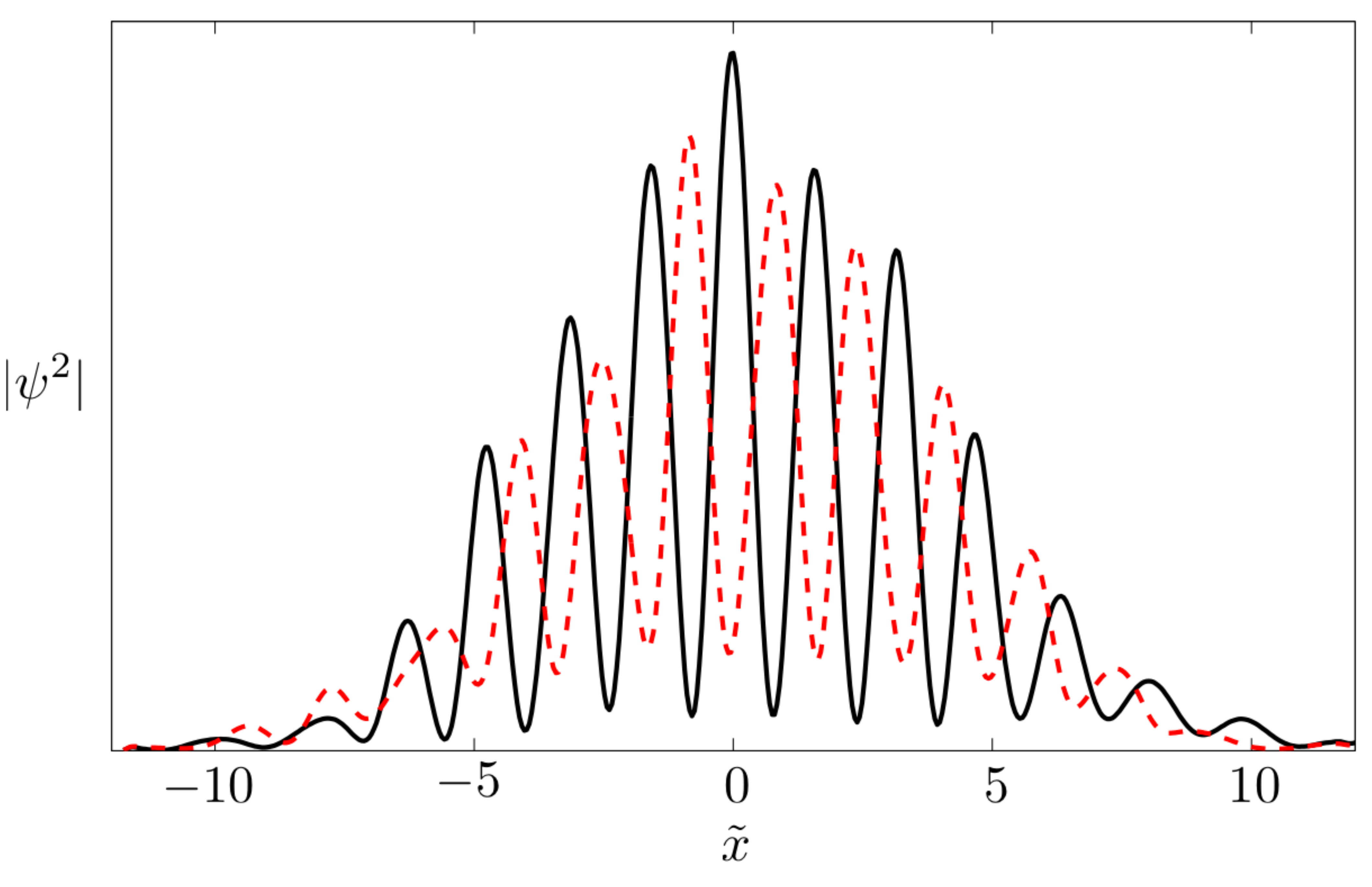,width=8.5cm}
\caption{The beamsplitter is removed from the interferometer and the bi-localized system is allowed to recombine. Due to the coherent superposition an interference pattern is observed for $F_Q=2.0023$ (red dashed line) and $F_Q=3.9998$ (black solid line). The difference in fringe contrast is apparent.}
\label{fringes}
\end{figure}

Evidently the scheme presented here has an immediate experimental appeal as many of its components are readily implementable. When a coherent bilocalised state is created the detection of this state can be achieved by measuring the fringe visibility of the two particle interference which is maximal in the presence of a NOON state  \cite{castin}. This can be done by exploiting the free oscillations in the harmonic trap after removing the barrier and a simulation of these fringes is shown in Fig.~\ref{fringes}. The solid line shows the pattern associated with the generated NOON state for $F_Q=3.9998$ at $g=-7$ and $\kappa=0.4$ and the dashed line shows what one would obtain for a state near the shot noise limit, $F_Q=2.0023$ at $g=-7$ and $\kappa=1.0985$. The difference in fringe contrast near the shot noise limit and near the Heisenberg limit can be clearly seen. One could also implement the scheme described in \cite{gooldheaney} where after the two sides of the trap are allowed to interfere at a beamsplitter the correlations are measured by counting the atoms collected at two different detectors. 

\section{Conclusions and Discussions}
\label{conclusions}
We have presented a comprehensive analysis of two interacting particles in a harmonic oscillator interferometer. By considering a wide range of parameters we have demonstrated the importance of the inter-particle interaction and its necessity in creating metrologically useful states. By employing exact numerical diagonalization methods we were able to study the type of states dynamically created and assess their value by studying the correlations via the Quantum Fisher Information and von Neumann entropy. The QFI is a useful metric for determining a states use in metrology and we found that the maximally achievable values depends strongly on the number of scattering events. After scattering on the barrier once, the attractively interacting particles were able to exceed the shot noise limit and even create NOON states for certain parameters. However, for repulsively interacting particles a single beam-splitting process does not allow to exceed this limit. After a second scattering from the barrier, thus realizing a Michelson interferometer, we found that NOON states could be created for both kinds of interactions, even though the range of potential parameters in the repulsive case was more limited.  As previously noted, although our study explicitly considers a delta-function barrier, the same results hold if one replaces it with a Gaussian barrier of finite width. In this instance the exact values of interaction strength and barrier height for optimal state generation will be slightly different to those found here, however the qualitative conclusions remain unaffected.

Let us finally comment extending the presented results to larger particle numbers. Treating the two-particle system has allowed us to rigorously assess the effect the inter-particle interaction has on the generation of metrologically useful states, while also allowing us to explore the entanglement dynamics via the von Neumann entropy. Going beyond two particles is computationally extremely costly, and it is clear from our study that  for ensembles of repulsive atoms highly correlated states will be very sensitive to the parameters involved. In fact, first calculations in the TG regime have shown that the generation of NOON states cannot be achieved this way. However for attractive interactions, even for a larger number of particles in the system, the scheme presented here should realize NOON states. As we have shown, when the atoms are strongly attractive, the bonds formed between the atoms are extremely hard to break, hence after the scattering processes they are much more likely to remain spatially close. This is somewhat analogous to the behaviour of bright solitons recently analyzed~\cite{gardiner,martin}.

\acknowledgments 
The authors would like to thank John Goold, Mauro Paternostro, Simon Gardiner and Dave Rea for inspiring and fruitful discussions, also thanks to Mark Kennedy and Lee O' Riordan for invaluable technical assistance. This work was supported by Science Foundation of Ireland under Project No. 10/IN.1/I2979 and by the Irish Research Council through the Embark Initiative RS/2009/1082.


\begin{thebibliography}{99}
\bibitem{nature}T. Schumm, S. Hofferberth, L. M. Andersson, S. Wildermuth, S. Groth, I. Bar-Joseph, J. Schmiedmayer, and P. Kr{\"u}ger, Nature Phys. {\bf 1}, 57 (2005).

\bibitem{atomoptical} Y. J. Wang, D. Z. Anderson, V. M. Bright, E. A. Cornell, Q. Diot, T. Kishimoto, M. Prentiss, R. A. Saravanan, S. R. Segal, and S. Wu, Phys. Rev. Lett. {\bf 94}, 090405 (2005).

\bibitem{BECinter1} Y. Shin, M. Saba, T. A. Pasquini, W. Ketterle, D. E. Pritchard, and A. E. Leanhardt, Phys. Rev. Lett. {\bf 92}, 050405 (2004).

\bibitem{bana} K. Banaszek, R. Demkowicz-Dobrzanski and I. A. Walmsley, Nature Photonics {\bf 3}, 673 - 676 (2009). 

\bibitem{mitchell} M. W. Mitchell, J. S. Lundeen and A.M. Steinberg, Nature {\bf 429}, 161-164 (2004).

\bibitem{walther} P. Walther et al, Nature {\bf 429}, 158-161 (2004).

\bibitem{rarity} J. G. Rarity et al, Phys. Rev. Lett. {\bf 65}, 1348 (1990).

\bibitem{afek} I. Afek, O. Ambar and Y. Silberberg, Science {\bf 328}, 879 (2010).

\bibitem{BECinter2} F. Baumg\"artner, R. J. Sewell, S. Eriksson, I. Llorente-Garcia, J. Dingjan, J. P. Cotter, and E. A. Hinds, Phys. Rev. Lett. {\bf 105}, 243003 (2010).

\bibitem{sackett08} J. H. T. Burke, B. Deissler, K. J. Hughes, and C. A. Sackett, Phys. Rev. A. {\bf 78}, 023619 (2008).

\bibitem{weiss12} B. Gertjerenken, T. P. Billam, L. Khaykovich, and C. Weiss, Phys. Rev. A {\bf 86}, 033608 (2012).

\bibitem{gardiner} J. L. Helm, T. P. Billam, and S. A. Gardiner, Phys. Rev. A. {\bf 85}, 053621 (2012)

\bibitem{castin} C. Weiss and Y. Castin, Phys. Rev. Lett. {\bf 102}, 010403 (2009).

\bibitem{streltsov} A. I. Streltsov, O. E. Alon, and L. S. Cederbaum, Phys. Rev. A. {\bf 80}, 043616 (2009).

\bibitem{sackett} R. H. Leonard and C. A. Sackett, Phys. Rev. A. {\bf 86}, 043613 (2012).

\bibitem{nakagawa} M. Horikoshi and K. Nakagawa, Phys. Rev. Lett. {\bf 99}, 180401 (2007).

\bibitem{zozulya} R. P. Kafle, D. Z. Anderson and A. A. Zozulya, Phys. Rev. A. {\bf 84}, 033639 (2011).

\bibitem{martin} A.D. Martin and J. Ruostekoski, New J. Phys. {\bf 14}, 043040 (2012). 

\bibitem{dunningham} J. J. Cooper, D. W. Hallwood, J. A. Dunningham and J. Brand, Phys. Rev. Lett. {\bf 108}, 130402 (2012); J. A. Dunningham, J. J. Cooper and D. W. Hallwood, arXiv:1102.0164, To Appear AIP Proceedings (2012).

\bibitem{braunstein} S. L. Braunstein and C. M. Caves, Phys. Rev. Lett. {\bf 72}, 3439 (1994).

\bibitem{Olshanni} M. Olshanii, Phys. Rev. Lett. {\bf 81}, 938 (1998).

\bibitem{julienne} C. Chin, R. Grimm and P. Julienne, Rev. Mod. Phys. {\bf 82}, 1225 (2010).  

\bibitem{bus98} Th. Busch, B. G. Englert, K Rzazewski  and M Wilkens, Found. Phys. {\bf 28}, 549 (1998).

\bibitem{DVR1} D. Baye and P. H. Heenen, J. Phys. A: Math. Gen. {\bf 19}, 2041 (1986).
\bibitem{DVR2} J. Light and J. T. Carrington, Adv. Chem. Phys. {\bf 114}, 263 (2000).
\bibitem{fisher} C. W. Helstrom, {\it Quantum Detection and Estimation Theory}, Academic Press, New York (1976).

\bibitem{gooldheaney} J. Goold, Libby Heaney, Th. Busch, and V. Vedral, Phys. Rev. A {\bf 80}, 022338 (2009).

\bibitem{2photon} Florian Wolfgramm, Chiara Vitelli, Federica A. Beduini, Nicolas Godbout, and Morgan W. Mitchell, Nature Photonics {\bf 7}, 28Ð32 (2013).

\bibitem{gooldtonks} D. S. Murphy, J. F. McCann, J. Goold, and Th. Busch, Phys. Rev. A. {\bf 76}, 053616 (2007).



\end{thebibliography}
\end{document}